\documentclass{svjour3}
\usepackage[shortlabels]{enumitem}
\smartqed  
\usepackage{graphicx}
\usepackage{bm}
\usepackage[margin=1in]{geometry}
\usepackage{booktabs}
\usepackage[linesnumbered,ruled,vlined]{algorithm2e}

\SetCommentSty{mycommfont}
\usepackage{algorithmic}
\usepackage{xcolor}
\SetKwInput{KwData}{Data}      
\SetKwInput{KwResult}{Result}  
\SetKwSty{textbf}             
\SetKwFor{For}{for}{do}{end}   
\SetKwComment{Comment}{/* }{ */} 
\DontPrintSemicolon             
\SetKwFunction{Solve}{Solve}     
\SetKw{KwIn}{in}               
\SetKw{KwUsing}{using}         
\SetKw{KwFrom}{from}           
\SetKw{KwWith}{with}         
\RequirePackage{fix-cm}
\usepackage{natbib}
\usepackage{xcolor}
\usepackage{amssymb}
\usepackage{amsmath}
\usepackage{lineno}
\usepackage{setspace}
\usepackage{hyperref}
\usepackage{longtable}
\usepackage{booktabs,subcaption,amsfonts,dcolumn}
\usepackage[compatibility=false]{caption}
\usepackage{algorithmic}
\usepackage{pifont}
\usepackage{siunitx}
\usepackage{booktabs,colortbl,array}
\usepackage{pgfplotstable}
\pgfplotsset{compat=1.8}
\usepackage{multirow}
\usepackage{hhline}
\usepackage{appendix}
\usepackage{enumitem}
\usepackage{booktabs}
\usepackage{tabularx}
\usepackage{cases}
\usepackage{empheq}
\usepackage{lmodern}

\definecolor{rulecolor}{RGB}{0,71,171}
\definecolor{tableheadcolor}{gray}{0.92}

\author{Milad Panahi}

\journalname{Submitted to }

\begin{document}
\title{\bfseries Probabilistic Inverse Modeling of Contaminant Transport via a Conditioned-on-Design Bayesian Physics Informed Neural Network}
\titlerunning{A Conditioned-on-Design Bayesian Physics Informed Neural Network (CoDe-BPINN)}
\author{Milad Panahi \and
        Giovanni Michele Porta * \and
         Monica Riva \and
        Alberto Guadagnini}
\date{Received: date / Accepted: date}
\institute{M. Panahi (0000-0002-8776-5297)\and G. M. Porta* (0000-0002-0636-373X) \and A. Guadagnini (0000-0003-3959-9690) \and M. Riva (0000-0002-7304-4114) \at Dipartimento di Ingegneria Civile e Ambientale, Politecnico di Milano, Piazza L. da Vinci 32, 20133 Milano, Italy \\
    * To whom correspondence should be addressed: \email{giovanni.porta@polimi.it}
}

\maketitle

\begin{abstract}
We address the inverse problem of reactive transport in heterogeneous porous media, where unknown model parameters must be inferred from sparse experimental observations. The problem is complicated by strong nonlinearities, spatial heterogeneity, and limited data availability. We propose a Conditioned-on-Design Bayesian Physics-Informed Neural Network (CoDe-BPINN), which combines a domain-decomposed PINN solver with a Bayesian inference network that learns the conditional distribution of model parameters given experimental design variables. The framework is trained by maximizing a physics-informed Evidence Lower Bound (ELBO), enabling simultaneous reconstruction of spatiotemporal concentration fields, probabilistic parameter estimation, and uncertainty quantification. We demonstrate the approach using laboratory experiments on contaminant transport through a multilayer porous column with an iodinated contrast medium. The model accurately reproduces breakthrough dynamics while revealing systematic parameter dependence on flow conditions. In particular, it identifies a nonlinear decrease in effective sorption capacity with increasing flow rate and porosity, consistent with kinetic limitations and reduced adsorbent mass. The Bayesian formulation also uncovers a strong negative correlation between sorption affinity and sorption capacity, quantifying the intrinsic non-identifiability of the inverse problem. CoDe-BPINN provides a robust framework for parameter inference and uncertainty quantification in data-scarce reactive transport problems.

\section*{Article Highlights}
\begin{itemize}
    \item[\ding{70}] CoDe-BPINN unifies Bayesian PINNs with conditional multivariate parameter inference.
    \item[\ding{70}] Bayesian inference quantifies parameter non-identifiability through posterior correlations.
    \item[\ding{70}] The model learns design-dependent parameter relationships from sparse noisy observations.
\end{itemize}


\end{abstract}

\section{Introduction}
\label{sec:ch4_introduction}

Management of water resources is increasingly challenged by proliferation of contaminants of emerging concern (CECs) in the global water cycle. Consequently, development of advanced remediation technologies, such as adsorption onto engineered porous media (e.g., organo-clays), is becoming a priority in environmental engineering scenarios \citep{patel2022comparison}. Modeling bottlenecks resulting in a broadly documented inability to provide accurate and unambiguous quantification of reactive transport dynamics under transient flow conditions still undermines effective transition from material synthesis to process-scale implementation. Some of the key challenges in this context are related to our non-exhaustive knowledge of the porous media across which they take place and the non-linear and scale-dependent nature of physicochemical parameters governing their evolution therein.

A common approach to describe the fate of contaminants in porous media relies on the Advection-Dispersion-Reaction (ADR) partial differential equations (PDEs) \citep{rubin2003applied}. While the mathematical structure of these conservation laws is well-known, the constitutive parameters embedded therein (most notably the dispersion tensor and reaction equilibr kinetic coefficients) are often unknown, spatially heterogeneous, and scale-dependent \citep{bear2013dynamics}. Traditional workflows for model parameter estimation typically rely on batch equilibrium experiments to determine sorption isotherms (e.g., Langmuir or Freundlich) \citep{foo2010insights}. However, it is well-documented that parameters derived from static batch tests are often not representative of dynamic behaviors observed in laboratory column experiments, mainly due to hydrodynamic effects inherent to flow systems \citep{limousin2007sorption, worch2008fixed}. Hence, model calibration is more appropriately grounded on data such as, e.g., dynamic breakthrough curves (BTCs) obtained from column experiments.

This inverse problem, namely the inference of spatially distributed or state-dependent parameters from sparse (and noisy) observations of system states across the domain and/or at its boundaries, is notoriously ill-posed \citep{tarantola2005inverse}. Classical gradient-based approaches, such as the Levenberg-Marquardt algorithm or adjoint-based methods, are deterministic optimization techniques designed to efficiently minimize a prescribed objective function. While computationally effective, they do not resolve the intrinsic ill-posedness of the inverse problem and often converge to a single (possibly local) solution among many plausible alternatives. As a result, they frequently suffer from non-uniqueness (equifinality), whereby multiple parameter sets produce indistinguishable model responses \citep{beven2001equifinality, carrera2005inverse}. Furthermore, while these methods can provide approximate parameter confidence intervals through local linearization (e.g., via error covariance matrices), they often fail to fully characterize their full (posterior) probability distribution or to capture structural (epistemic) uncertainty in strongly non-linear systems. These elements are notably relevant in the context of flow and contaminant transport, as the quantification of uncertainty is as critical as predictive capability itself in defining the reliability of modeling approaches \citep{gupta2006model, dell2019solute, tartakovsky2013assessment,  scheidt2018quantifying}. More specifically, experimental observations or reactive transport processes typically embed information at selected space-time locations (e.g., through breakthrough curves), which restricts the information available for parameter estimation \cite{taylor2026global,ceriotti2019double}. 

In recent years, Scientific Machine Learning (SciML) has emerged as a transformative paradigm to bridge the gap between observational data and formulation of physical laws \citep{lavin2021simulation}. Physics-Informed Neural Networks (PINNs) constitute a technique of emerging interest \citep{raissi2019physics, raissi2024physics, luo2025physics}, which allows embedding the residuals of the PDE governing the system dynamics directly into the neural network loss function. PINNs serve as universal function approximators \citep{lagaris1998artificial} that are constrained by given mathematical formulations of physical laws and are designed to to learn solutions (i.e., system state variables of interest) from sparse data. PINNs provide a novel approach to numerical approximation that allows circumventing some limitations of classical methods such as, e.g., mesh-dependency. While standard PINNs have demonstrated remarkable success in solving inverse problems for parameter identification \citep{kovachki2021neural, lu2021learning, li2020fourier, chen2020physics}, they typically provide deterministic point values valid only for a specific training scenario \citep{raissi2023open}. Hence, they do not naturally provide confidence intervals for the estimated parameters, nor quantify predictive uncertainty associated with the reconstructed solution. To address these elements, Bayesian Physics-Informed Neural Networks (BPINNs) have been proposed \citep{yang2021b, meng2021multi, sun2020physics}. By placing probability distributions over network weights and treating physical model parameters as random variables, BPINNs enable one to distinguish between uncertainty in the neural approximation (approximation uncertainty) and in the identification of model parameters (parameter uncertainty) \citep{psaros2023uncertainty, lee2009comparative, Maina2018}.

Despite these advancements, significant challenges still remain to the application of BPINNs to multi-physics problems in heterogeneous media. Standard global neural networks often suffer from spectral bias \citep{rahaman2019spectral, wang2022and} and optimization pathologies \citep{fletcher2000practical} when the solution contains sharp gradients, such as those found at material interfaces in multi-layered adsorption columns \citep{wang2021understanding}. Domain decomposition methods, such as the Conservative PINN (CPINN) \citep{jagtap2020conservative}, have then been proposed to mitigate this limitation by employing separate sub-networks for diverse physical subdomains. Yet, their integration into a fully Bayesian inverse modeling framework remains underexplored.

This study constitutes an additional step of a comprehensive research framework designed to address parametric uncertainty in flow and reactive transport modeling in porous media leveraging a PINN framework. In a first study \citep{panahi2025modeling}, we developed PINNs specifically tailored to Uncertainty Quantification (hereafter termed \textit{PINN-UU}) to address forward propagation of parametric uncertainty. Consistent with broader developments in the field \citep[e.g.,][]{flores2025improved, sun2020surrogate, zhu2019physics, CHEN2021110666, asher2015review} we demonstrated that neural networks can serve as efficient surrogates for mapping stochastic inputs to system. In a subsequent phase, we tackled the challenge of spatial heterogeneity upon developing differentiable solvers \citep{panahi2026pids} that rely on automatic differentiation techniques \citep{thuerey2021physics, NEURIPS2018_DP, innes2019differentiable} to learn continuous solution manifolds in complex, heterogeneous hydraulic conductivity fields without the need for retraining. While these works established the foundation for parameterized forward modeling, they rest on the assumption that the underlying physical laws and parameters are deterministically known or statistically characterized.

Here, we address both forward and inverse modeling in a probabilistic setting. We consider transport of iodinated contrast media (ICM), such as, e.g., iohexol and iopamidol. ICMs constitute a distinct class of persistent micropollutants whose chemical stability and recalcitrance raise significant environmental and toxicological concerns. These compounds are widely used in medical imaging and are characterized by high polarity, hydrophilicity, and remarkable chemical stability, rendering them recalcitrant to conventional biological wastewater treatment processes \citep{KHAN2024116506, sengar2021occurrence}. Their ubiquitous presence in surface water, groundwater, and even drinking water sources poses potential risks, particularly regarding potential formation of toxic iodinated disinfection byproducts \citep{kormos2011occurrence, duirk2011formation}. We apply this framework to the transport of the ICM Iohexol in a stratified sand-clay (laboratory-scale) column, as described in \cite{khan2026removal}. The experimental setup is designed to remove Iohexol via sorption to an engineered clay material. The effective interface properties are thus critical to characterize the performance of the system. Here, the effective sorption properties are hypothesized to be a dynamic function of the experimental design rather than a static constant. Framing the analysis according to the perspective of \citet{SAdelloca}, we explicitly distinguish between \textit{operational variables} (i.e., controllable operational design parameters such as fluid flow rate and inlet concentration) and \textit{stochastic model parameters} (i.e., uncertain quantities such as sorption affinity and capacity).

Our main objective is to enable the model to autonomously discover state-dependent functional relationships, specifically characterizing the decline of the effective sorption capacity of the porous medium under diverse flow regimes while rigorously quantifying parametric uncertainty. To achieve this, we propose a probabilistic model, formulated as a Conditioned-on-Design Bayesian Physics Informed Neural Network (CoDe-BPINN). This architecture integrates a domain-decomposed Bayesian solver (based on the CPINN philosophy) with a Multivariate Parameter Inference Network, in a conditioning HyperNetwork style \citep{ha2016hypernetworks, CHEN2022110996, nichol2018reptile}. By assuming physical parameters as random variables conditioned to operational variables we transform the inverse problem into a \textit{conditional density estimation} task. This formulation allows us to disentangle the variability arising from operational choices from the lack of knowledge regarding physical properties of the system, offering a robust basis for process monitoring and experimental design.

Unlike standard Bayesian inversion, which typically infers a posterior distribution of otherwise unknown model parameters for a single dataset, our framework learns the \textit{conditional probability distribution} of model parameters as a function of the operational design space, conditioned on the sparse observational data. By minimizing a physics-informed Evidence Lower Bound (ELBO), the model aims at achieving the following:

\begin{enumerate}[label=\roman*.]
    \item \textbf{Simultaneous Reconstruction and Inference:} the model yields a reconstruction of the full spatiotemporal concentration field while simultaneously inferring the governing physical parameters, thus effectively solving the coupled forward-inverse problem in a single differentiable loop.
    
    \item \textbf{Understanding Learned Relationships between Physical Parameters:} rather than optimizing static scalar values, the model learns a continuous mapping from operational variables (e.g., flow rate, inlet solute concentration, or porosity of the domain) to physical coefficients. This allows for the autonomous discovery of non-linear relationships and the influence of experimental design on estimated properties.
    
    \item \textbf{Quantification of Structural Identifiability:} by utilizing a multivariate formulation, the framework explicitly captures the correlation structure between coupled parameters (e.g., the trade-off between affinity and capacity). This identifies regimes of practical non-identifiability (i.e., equifinality) that deterministic methods fail to detect.
\end{enumerate}

The remainder of the paper is organized as follows. Section~\ref{sec:experimental_setup} provides a summary of the experimental setting and dataset that we consider in our study. Section~\ref{sec:ch4_methodology} details the mathematical formulation of the parameterized Advection-Dispersion-Reaction system and our proposed differentiable Bayesian framework. Section~\ref{sec:ch4_results} illustrates and discusses the ensuing numerical results, including validation against high-fidelity solvers and the analysis of the discovered functional relationships between model parameters. Finally, Section~\ref{sec:ch4_conclusion} provides key conclusions and outlines future research directions.

\begin{table}[!ht]
    \centering
    \caption{Summary of the main characteristics of the 9 experimental configurations considered in the study. Here, $q$ denotes Darcy flux [mL/min]; $C_0$ is Inlet Concentration; and $\varepsilon_{\Omega_2}$ is porosity of the reactive middle layer of the column (Fig. \ref{fig:exp_setup}).}
    \label{tab:exp_matrix}
    \begin{tabularx}{\textwidth}{c c c c >{\centering\arraybackslash}X} 
    \toprule
    \textbf{Exp ID} & \textbf{Darcy Flux} $q$ & \textbf{Inlet Concentration} $C_0$ & \textbf{Porosity} $\varepsilon_{\Omega_2}$ & \textbf{Variable Tested} \\
     & [mL/min] & [mg/L] & [-] & \\
    \midrule
    1 & 1.99 & 100 & 0.371 & Baseline \\
    2 & 4.97 & 100 & 0.371 & Flow Rate \\
    3 & 12.43 & 100 & 0.371 & Flow Rate \\
    \midrule
    4 & 1.99 & 150 & 0.371 & Concentration \\
    5 & 1.99 & 200 & 0.371 & Concentration \\
    6 & 1.99 & 250 & 0.371 & Concentration \\
    7 & 1.99 & 300 & 0.371 & Concentration \\
    \midrule
    8 & 1.99 & 100 & 0.424 & Porosity (Low Clay) \\
    9 & 1.99 & 100 & 0.399 & Porosity (Med Clay) \\
    \bottomrule
    \end{tabularx}
\end{table}

\section{Experimental Setup and Data Acquisition}
\label{sec:experimental_setup}

Data employed in this study are derived from a comprehensive experimental campaign detailed in \citet{khan2026removal} which is aimed at investigating removal of Iodinated Contrast Media (ICM) agents using engineered clay adsorbents. We focus on transport of Iohexol through a fixed-bed (laboratory-scale) column packed with a composite of quartz sand and modified montmorillonite organo-clay (Mt-II), as described in the following.

\subsection{Materials and Column Configuration} \label{exp_geometric}
The adsorbent medium consists of a dually-modified montmorillonite (Mt-II), synthesized via sequential intercalation of cationic (CTPC) and anionic (SDS) surfactants to enhance affinity for polar organic micropollutants. 

\begin{figure}[!ht]
    \centering
    \includegraphics[width=0.5\textwidth]{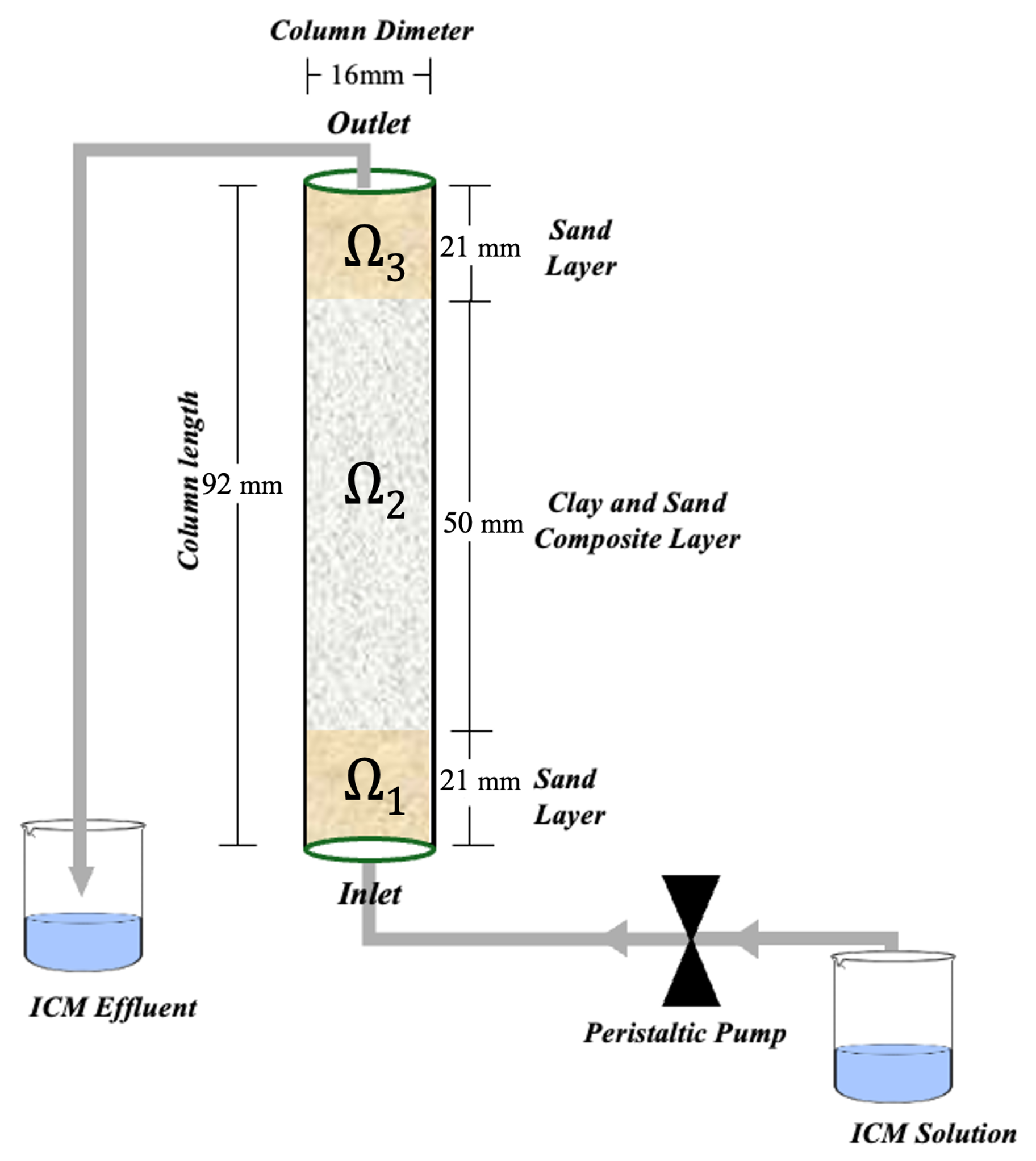}
    \caption{Schematic of the three-layer fixed-bed column experiment. A solution containing the ICM Iohexol is injected in the column. Iohexol concentration data are collected at discrete observation times from the effluent at the column outlet.}
    \label{fig:exp_setup}
\end{figure}

As illustrated in Figure~\ref{fig:exp_setup}, the fixed-bed column domain $\Omega = [0, L]$ (with total length of $L=92$ mm, an internal diameter of $d = 16$ mm, and a cross-sectional area $A \approx 201.06$ mm$^2$) is stratified into the following three zones to ensure hydraulic stability and prevent washout of fine clay particles:
\begin{enumerate}[label=\roman*.]
    \item \textbf{Inlet Zone ($\Omega_1$):} A 21 mm layer of pure, washed quartz sand (50-70 mesh) acting as a flow distributor.
    \item \textbf{Reactive Zone ($\Omega_2$):} A 50 mm central layer comprising the Mt-II organo-clay dispersed within the sand matrix. This is the sole domain where reactive transport (sorption) occurs. The porosity of this layer, $ \varepsilon_{\Omega_2}$ varies depending on the clay packing density.
    \item \textbf{Outlet Zone ($\Omega_3$):} A 21 mm layer of pure quartz sand preventing adsorbent migration and stabilizing the outflow.
\end{enumerate}

\subsection{Batch vs. Column Dynamics}
Prior to column studies, batch equilibrium experiments were performed to characterize the intrinsic sorption properties of the Mt-II clay. The ensuing equilibrium data could be interpreted through the Langmuir isotherm formulation:
\begin{equation}
q_e = \frac{\beta \alpha C_{eq}}{1 + \alpha C_{eq}}
\label{eq:langmuir_batch}
\end{equation}
where $q_e$ [mg/g] and $C_{eq}$ [mg/L] denote the surface- and liquid-phase concentrations of Iohexols, respectively; $\beta$ [mg/g] is the maximum adsorption capacity of the engineered porous medium, and $\alpha$ [L/mg] is the Langmuir affinity constant.

Parameters associated with batch experiments (i.e., $\beta$ and $\alpha$) represent thermodynamic equilibrium conditions (i.e., for time $t \to \infty$). In dynamic column systems, hydrodynamic dispersion and kinetic limitations (non-equilibrium transport) often result in effective parameters that differ from their batch-grounded counterparts. We treat the batch-derived values ($\beta_{batch} \approx 60$ mg/g, $\alpha_{batch} \approx 0.03$ L/mg) as \textit{upper bound priors} for our inverse modeling framework, allowing the CoDe-BPINN to discover the effective model parameters $\boldsymbol{\theta}_{col}$ that are consistent with the hydrodynamic constraints emerging during the experiments.

\subsection{Experimental Design and Observational Data}
\label{sec:ch4_training_data}
To evaluate transport of Iohexol under a broad range of conditions, a suite of 9 column experiments was performed (see Table~\ref{tab:exp_matrix}). These experiments were designed to assess the response of the system to three key operational variables:
\begin{enumerate}[label=\roman*.]
    \item \textbf{Inlet Concentration ($C_0$)}, enabling one to investigate the saturation of the adsorbent media (Experiments 1, 4, 5, 6, and 7);
    \item \textbf{Darcy Flux ($q$)}, to probe hydrodynamic residence time and dispersive effects (Experiments 1, 2, and 3);
    \item \textbf{Middle Layer Porosity ($\varepsilon_{\Omega_2}$)}, which serves as a proxy for the clay volume fraction (adsorbent mass) within the reactive zone (Experiments 1, 8, and 9).
\end{enumerate}
The specific combinations for these 9 experiments are detailed in Table~\ref{tab:exp_matrix}. The resulting dataset (hereafter denoted as $\mathcal{D}_{obs}$) is composed by the temporal history of concentration recorded at the outlet (also defined as breakthrough curves) for each case. 

\section{Methodology: A Differentiable Physics-Informed Learning Framework}
\label{sec:ch4_methodology}

We propose a unified computational framework for solving a coupled forward-inverse problem under parametric uncertainty. The architecture integrates a domain-decomposed Bayesian Neural Network (BNN) solver, denoted as $\mathcal{S}$, with a multivariate parameter inference network, $\mathcal{N}_{inv}$, into a single differentiable computational graph. This enables simultaneous adherence to sparse observational data and physics-constrained optimization via gradient-based methods.

\subsection{General Framework for Mixed-Deterministic Parameterized PDEs} \label{sec:ch4_General_Framework}
Let the spatial domain be an open set $\Omega \subset \mathbb{R}^d$. We assume that the system of interest can be modelled via a PDE that is characterized by a vector of operational parameters $\boldsymbol{\lambda} \in \Lambda \subset \mathbb{R}^{d_\lambda}$ and a vector of unknown physical parameters $\boldsymbol{\theta} \in \Theta \subset \mathbb{R}^{d_\theta}$. The relationship between parameters and the solution field $C(\mathbf{x}, \boldsymbol{\lambda}, \boldsymbol{\theta})$ is implicitly defined by a differential operator. We approximate the unknown $C(\mathbf{x}, \boldsymbol{\lambda}, \boldsymbol{\theta})$ using a domain-decomposed Bayesian neural network (the solver $\mathcal{S}$), parameterized by weights $\boldsymbol{\psi}$. At the same time, we approximate the inverse parameter map using a multivariate network $\mathcal{N}_{inv}$ characterized by internal parameters $\boldsymbol{\phi}$. We define a PDE residual operator $\mathcal{F}$ and a boundary residual operator $\mathcal{B}$. The parameterized PDE problem is to find a surrogate solution $\hat{C}$ for each instance $\boldsymbol{\lambda} \in \Lambda$ such that:
\begin{align}
\mathcal{F}[\mathbf{x}, \boldsymbol{\lambda}, \boldsymbol{\theta}; \hat{C}] &= 0, \quad \forall \mathbf{x} \in \Omega \\
\mathcal{B}[\mathbf{x}, \boldsymbol{\lambda}, \boldsymbol{\theta}; \hat{C}] &= 0, \quad \forall \mathbf{x} \in \partial\Omega
\end{align}
Within the taxonomy of Scientific Machine Learning, this represents a Mixed Deterministic PDE Problem \citep{zhang2019quantifying}. While the operator $\mathcal{F}$ is deterministic, but the full solution field is unknown, only sparse measurements ($\mathcal{D}_{obs}$) are available at the domain outlet. Furthermore, the physical parameters $\boldsymbol{\theta}$ are unknown and must be inferred. While intrinsic material properties are in principle independent of experimental design, we adopt a standard conceptual picture according to which the corresponding \textit{effective} values of these parameters might exhibit dependencies on the operational design parameters (i.e., $\boldsymbol{\theta} \approx g(\boldsymbol{\lambda})$). In our test case, this might be due to unresolved non-equilibrium dynamics or scaling effects leading to dependencies between surface reaction and transport in certain experimental regimes. Our primary goal is to simultaneously learn the forward solution map $\mathcal{G}: (\mathbf{x}, \boldsymbol{\lambda}) \mapsto C$ and the inverse map, which infers the parameter distribution consistent with the observations, i.e., $\mathcal{G}_{inv}: (\boldsymbol{\lambda}, \mathcal{D}_{obs}) \mapsto q_{\boldsymbol{\phi}}(\boldsymbol{\theta}|\boldsymbol{\lambda})$, where $q_{\boldsymbol{\phi}}(\boldsymbol{\theta}|\boldsymbol{\lambda})$ represents the posterior distribution of the physical parameters conditional to operational parameters.

\paragraph{Neural Network Solver.}
we introduce a dual-network architecture that approximates the unknown solution map and the inverse parameter map through parameterized neural networks, as depicted in Figure~\ref{fig:ch4_architecture}. A primary BPINN ($\mathcal{S}$), designed analogously to a CPINN \citep{JAGTAP2020113028} and composed of three sub-domain solvers $\{\mathcal{N}^{\Omega_1}, \mathcal{N}^{\Omega_2}, and \mathcal{N}^{\Omega_3}\}$, acts as an approximator of the unknown solution $C$. This network is coupled with a secondary network devoted to inverse modeling ($\mathcal{N}_{inv}$), which learns a probabilistic mapping from operational parameters to the latent physical parameters. In this context,
\begin{itemize}
    \item The \textbf{Forward Solution} $C(\mathbf{x}, \boldsymbol{\lambda})$ is approximated by a domain-decomposed Bayesian Neural Network (the Solver $\mathcal{S}$) with variational weights $\boldsymbol{\psi} = \{\boldsymbol{\psi}_1, \boldsymbol{\psi}_2, \boldsymbol{\psi}_3\}$ and yields a predictive distribution $\mathcal{P}(\hat{C} | \mathbf{x}, \boldsymbol{\lambda}; \boldsymbol{\psi})$ as output.
    \item The \textbf{Inverse Map} for $\boldsymbol{\theta}$ is approximated by a multivariate Parameter Network ($\mathcal{N}_{inv}$) with weights $\boldsymbol{\phi}$; as output, it yields the conditional statistics of the physical parameters, namely the mean vector $\boldsymbol{\mu}_{\ln\theta}(\boldsymbol{\lambda}; \boldsymbol{\phi})$ and the Cholesky factor of the corresponding covariance matrix $\mathbf{L}_{\ln\theta}(\boldsymbol{\lambda}; \boldsymbol{\phi})$.
\end{itemize}

The parameters of all networks $\Phi = \{\boldsymbol{\psi}, \boldsymbol{\phi}\}$ are optimized simultaneously by minimizing a composite physics-informed loss function that enforces the operators $\mathcal{F}$ and $\mathcal{B}$ on the network outputs and minimizes discrepancies with the sparse observational data $\mathcal{D}_{obs}$.

\paragraph{Coupled Training vs. Decoupled Inference.}
While the solver $\mathcal{S}$ and the Parameter Inference Network $\mathcal{N}_{inv}$ are jointly trained to ensure physical consistency, they operate as independent modules during inference. Once training is complete, information about physical parameters estimates $\boldsymbol{\theta}$ is implicitly encoded within the solver weights $\boldsymbol{\psi}$ across the operational design space $\boldsymbol{\lambda}$. Consequently, $\mathcal{S}$ serves as a standalone forward surrogate, enabling evaluation of the output variable without further queries to the inference network. At the same time, the trained $\mathcal{N}_{inv}$ can be independently queried to explicitly characterize the relationship between operational and physical parameters and, for instance, parameterize alternative numerical solvers, such as the Finite Difference Method (FDM) reference scheme employed in this study.

\subsection{Problem definition: The Parameterized Advection-Dispersion-Reaction System}\label{sec:ch4_problem_formulation}
We consider a physical system encompassing transient one-dimensional transport of the Iodinated Contrast Media (ICM) agent Iohexol across the saturated, stratified fixed-bed porous medium column described in Section~\ref{sec:experimental_setup}.

The system is characterized by the three operational design parameters introduced in Section~\ref{sec:ch4_training_data}. These are here used to characterize the system under diverse conditions and are grouped into the operational design vector $\boldsymbol{\lambda} = [\varepsilon_{\Omega_2}, q, C_0]^\top$. Unknown effective physical properties of the porous system are encapsulated in the stochastic parameter vector $\boldsymbol{\theta} = [a_{Ls}, a_{Lc}, \alpha, \beta]^\top$. These are defined as:
\begin{itemize}
    \item $a_{Ls}, a_{Lc}$: Longitudinal dispersivities for sand and clay layers [mm], respectively.
    \item $\alpha$: Langmuir sorption affinity constant [L/mg].
    \item $\beta$: Maximum sorption capacity [mg/g].
\end{itemize}

These parameters are piecewise constant in space, i.e. they are assumed constant within each layer, and are inferred by the network using available measurements $\mathcal{D}_{obs}$, collected for various combinations of the operational design parameters $\boldsymbol{\lambda}$.

The spatiotemporal evolution of solute concentrations system is described by the one-dimensional Advection-Dispersion-Reaction (ADR) equation. We formulate the problem in terms of a dimensionless concentration, $\hat{C}(\mathbf{x}, \boldsymbol{\lambda}) = C(x,t)/C_0$, where $\mathbf{x} = (x,t)$. The PDE operator, $\mathcal{F}$, is defined as:
\begin{equation}
\mathcal{F}[\hat{C}] := \frac{\partial \hat{C}}{\partial t} R_L(\mathbf{x}, \hat{C}; \boldsymbol{\lambda}, \boldsymbol{\theta}) - \frac{\partial}{\partial x} \left( D_L(\mathbf{x}; \boldsymbol{\lambda}, \boldsymbol{\theta}) \frac{\partial \hat{C}}{\partial x} \right) + v_x(x; \boldsymbol{\lambda}) \frac{\partial \hat{C}}{\partial x} = 0
\label{eq:ch4_strong_form}
\end{equation}
The pore water velocity is $v_x(x; \boldsymbol{\lambda})= q / \varepsilon(x)$, where $\varepsilon(x)$ is the local porosity. The dispersion coefficient is modeled as $D_L = a_L(x) |v_x| + D^*$ [mm$^2$/min], where $a_L(x)$ takes the value $a_{Lc}$ in $\Omega_2$ and $a_{Ls}$ elsewhere. The effective diffusion coefficient is defined as $D^* = 1.0 \times 10^{-6} \cdot (\varepsilon(x)^{4/3})$ [mm$^2$/min].

The non-linear Langmuir Retardation Factor, $R_L$, is active only in the reactive layer ($\Omega_2$) and is defined as:
\begin{equation}
R_L(\mathbf{x}, \hat{C}; \boldsymbol{\lambda}, \boldsymbol{\theta}) = 1 + \mathbb{I}_{\Omega_2}(x) \frac{\rho_b}{\varepsilon(x)} \left( \frac{\alpha \beta}{(1 + \alpha C_0 \hat{C})^2} \right)
\label{eq:ch4_retardation}
\end{equation}
where $\mathbb{I}_{\Omega_2}$ is the indicator function restricting sorption to the clay layer $\Omega_2$, and $\rho_b$ is the bulk density [mg/mm$^3$]. Note that the formulation explicitly retains the inlet concentration $C_0$ [mg/L] in the denominator, capturing the concentration-dependent non-linearity of the Langmuir equilibrium adsorption model \citep{fetter1999contaminant}.

The system is subject to the boundary operator $\mathcal{B}$, with the following boundary and initial conditions:
\begin{subnumcases}{\label{eq:ch4_bcs}}
\hat{C}(x, t=0) = 0, & $\forall x \in \Omega \quad \text{(Initial Condition)}$ \\
\hat{C}(x=0, t) = 1, & $\forall t > 0 \quad \text{(Inlet Dirichlet BC)}$ \\
\frac{\partial \hat{C}}{\partial x}(x=L, t) = 0 & $\forall t > 0 \quad \text{(Outlet Neumann BC)}$
\end{subnumcases}

\subsection{Bayesian Physics Informed Neural Network (BPINN) Formulation}
\label{sec:bayesian_formulation}
As stated in Section \ref{sec:ch4_methodology}, we adopt here a Bayesian approach. In contrast to deterministic PINNs, where the network weights are fixed after training and represent single (point) estimates, we instead model both the solver weights $\boldsymbol{\psi}$ and the physical parameters $\boldsymbol{\theta}$ as random variables, thereby explicitly accounting for uncertainty in their values.

Our goal is to compute the joint posterior distribution of these quantities, conditioned on the observational data $\mathcal{D}_{obs}$ and the governing physical laws, for a given set of operational design parameters $\boldsymbol{\lambda}$. The physical constraints are defined by the operators $\mathcal{F}[\hat{C}; \boldsymbol{\lambda}, \boldsymbol{\theta}]=0$ and $\mathcal{B}[\hat{C}; \boldsymbol{\lambda}, \boldsymbol{\theta}]=0$. Using Bayes' theorem, the full (joint) posterior is:
\begin{equation}
p(\boldsymbol{\psi}, \boldsymbol{\theta} | \mathcal{D}_{obs}, \boldsymbol{\lambda}) \propto p(\mathcal{D}_{obs} | \boldsymbol{\psi}, \boldsymbol{\theta}, \boldsymbol{\lambda}) p(\boldsymbol{\psi}, \boldsymbol{\theta} | \boldsymbol{\lambda})
\label{eq:bayes_simple}
\end{equation}
where $p(\mathcal{D}_{obs} | \boldsymbol{\psi}, \boldsymbol{\theta}, \boldsymbol{\lambda})$ is the data likelihood. The prior, $p(\boldsymbol{\psi}, \boldsymbol{\theta} | \boldsymbol{\lambda})$, can be factored by assuming the solver weights $\boldsymbol{\psi}$ are independent of the operational design space $\boldsymbol{\lambda}$, yielding $p(\boldsymbol{\psi}, \boldsymbol{\theta} | \boldsymbol{\lambda}) = p(\boldsymbol{\psi}) p(\boldsymbol{\theta} | \boldsymbol{\lambda})$. Here, $p(\boldsymbol{\psi})$ is the prior distribution of the network weights, and $p(\boldsymbol{\theta} | \boldsymbol{\lambda})$ is the informative prior that reflects our assumptions of the way physical parameters depend on the operational design parameters. Following standard BPINN methodology, we incorporate the physical constraints as a "physics likelihood" term, $p(\mathcal{F}, \mathcal{B} | \boldsymbol{\psi}, \boldsymbol{\theta}, \boldsymbol{\lambda})$. While the form of this distribution is a mere modeling choice, we assume here for simplicity and tractability that the PDE residuals are distributed according to a zero-mean Gaussian, i.e., $p(\mathcal{F}|\cdot) \propto \exp(-||\mathcal{F}||^2)$:

\begin{equation}
p(\boldsymbol{\psi}, \boldsymbol{\theta} | \mathcal{D}_{obs}, \mathcal{F}, \mathcal{B}, \boldsymbol{\lambda}) \propto p(\mathcal{D}_{obs} | \boldsymbol{\psi}, \boldsymbol{\theta}, \boldsymbol{\lambda}) p(\mathcal{F}, \mathcal{B} | \boldsymbol{\psi}, \boldsymbol{\theta}, \boldsymbol{\lambda}) p(\boldsymbol{\psi}) p(\boldsymbol{\theta} | \boldsymbol{\lambda})
\label{eq:bayes}
\end{equation}
The exact computation of this posterior requires evaluating the model evidence, or \textbf{marginal likelihood}, $p(\mathcal{D}_{obs}, \mathcal{F}, \mathcal{B} | \boldsymbol{\lambda})$, which involves an intractable high-dimensional integral over $\boldsymbol{\psi}$ and $\boldsymbol{\theta}$. Hence, we employ Variational Inference (VI) to approximate the true posterior. We introduce a parameterized family of variational distributions, i.e. $q_{\boldsymbol{\psi}}(\boldsymbol{\psi})$ for the solver weights and $q_{\boldsymbol{\phi}}(\boldsymbol{\theta}|\boldsymbol{\lambda})$ for the physical parameters. The training objective is then to minimize the Kullback-Leibler divergence ($D_{KL}$) between this approximation and the true posterior, which is equivalent to maximizing the Evidence Lower Bound (ELBO):
\begin{equation}
\mathcal{L}_{ELBO} = \mathbb{E}_{q_{\boldsymbol{\psi}}q_{\boldsymbol{\phi}}} \left[ \ln p(\mathcal{D}_{obs} | \boldsymbol{\psi}, \boldsymbol{\theta}, \boldsymbol{\lambda}) + \ln p(\mathcal{F}, \mathcal{B} | \boldsymbol{\psi}, \boldsymbol{\theta}, \boldsymbol{\lambda}) \right] - D_{KL}[q_{\boldsymbol{\psi}}(\boldsymbol{\psi})||p(\boldsymbol{\psi})] - D_{KL}[q_{\boldsymbol{\phi}}(\boldsymbol{\theta}|\boldsymbol{\lambda})||p(\boldsymbol{\theta})]
\label{eq:elbo}
\end{equation}
The terms formulated as expectation quantify the quality of the reconstruction by capturing agreement with observed data as well as consistency with the governing physics, while the KL divergence terms act as regularizers. In practice, this objective is implemented as the composite trainable loss function detailed in Section~\ref{sec:ch4_composite_loss}, where expectations are approximated via Monte Carlo sampling. To enable gradient-based optimization, we adopt the reparameterization scheme proposed in\citep{kingma2013auto}, where stochastic variables are expressed as $w = \mu + \sigma \odot \epsilon$ with $\epsilon \sim N(0, I)$, thereby allowing backpropagation through the Bayesian network solver.

\subsection{From Likelihood-based formulations to a Trainable Loss Function}
\label{sec:physics_likelihood}
The ELBO in Eq.~\ref{eq:elbo} involves evaluating the log-likelihood of the physical constraints, $\ln p(\mathcal{F}, \mathcal{B} | \boldsymbol{\psi}, \boldsymbol{\theta}, \boldsymbol{\lambda})$. A key step in BPINNs is therefore to recast these constraints, originally expressed as deterministic equalities (e.g., $\mathcal{F}[\hat{C}; \boldsymbol{\lambda}, \boldsymbol{\theta}] = 0$), into a probabilistic framework. Following standard approaches \citep{yang2021b}, we treat the PDE residuals as noisy observations with zero mean and an associated precision (i.e., inverse variance).

Specifically, we assume the likelihood of the PDE residuals to be described through a Gaussian distribution:
\begin{equation}
p(\mathcal{F} | \boldsymbol{\psi}, \boldsymbol{\theta}, \boldsymbol{\lambda}) \propto \exp \left( - \frac{1}{2\sigma^2_{\mathcal{F}}} \sum_{j=1}^{N_f} ||\mathcal{F}[\hat{C}; \mathbf{x}_j, \boldsymbol{\lambda}, \boldsymbol{\theta}]||^2 \right)
\label{eq:physics_likelihood}
\end{equation}
Here, $\sigma^2_{\mathcal{F}}$ represents the variance associated with the residuals associated with the differential operator, thereby controlling the tolerance allowed for deviations from the governing equation. Maximizing the corresponding log-likelihood is equivalent to minimizing the Mean Squared Error (MSE) of the PDE residuals. Analogous likelihood terms are defined for the boundary ($\mathcal{B}$) and initial ($\mathcal{I}$) conditions, as well as, within our domain-decomposed architecture, for interface continuity constraints, which are formally defined in the following in Section~\ref{sec:ch4_composite_loss}.

The training objective involves evaluating the expected log-likelihood terms appearing in the ELBO (Eq.~\ref{eq:elbo}), where expectation is taken with respect to the joint variational posterior of the latent variables. Under the assumption that the solver weights and physical parameters are independent within the variational approximation, this joint density factorizes as $q_{\boldsymbol{\psi}}(\boldsymbol{\psi}) \cdot q_{\boldsymbol{\phi}}(\boldsymbol{\theta}|\boldsymbol{\lambda})$. Since an analytical evaluation of this high-dimensional expectation is unfeasible due to the non-linear structure of the neural networks, we approximate it via Monte Carlo sampling at each training iteration. Specifically, at each training iteration we sample the solver weights $\boldsymbol{\psi}$ from their variational posterior $q_{\boldsymbol{\psi}}(\boldsymbol{\psi})$ and the physical parameters $\boldsymbol{\theta}$ from the \textbf{Parameter Inference Network} ($\mathcal{N}_{inv}$). The latter represents the variational distribution $q_{\boldsymbol{\phi}}(\boldsymbol{\theta}|\boldsymbol{\lambda})$ by outputting the mean $\boldsymbol{\mu}_{\ln\theta}$ and the Cholesky factor $\mathbf{L}_{\ln\theta}$ of the log-parameter covariance matrix, as introduced in Section~\ref{sec:ch4_General_Framework}.
This sampling procedure converts the ELBO from a theoretical formulation into a tractable stochastic objective. In this setting, the expected log-likelihood terms naturally correspond to the MSE loss terms typical of deterministic PINNs, scaled (i.e., weighted) by their corresponding inverse variances. This directly leads to the composite loss function described in Section~\ref{sec:ch4_composite_loss}, yielding a computationally tractable framework for training the CoDe-BPINN.

\begin{figure}[!ht]
    \centering
    \includegraphics[width=\textwidth]{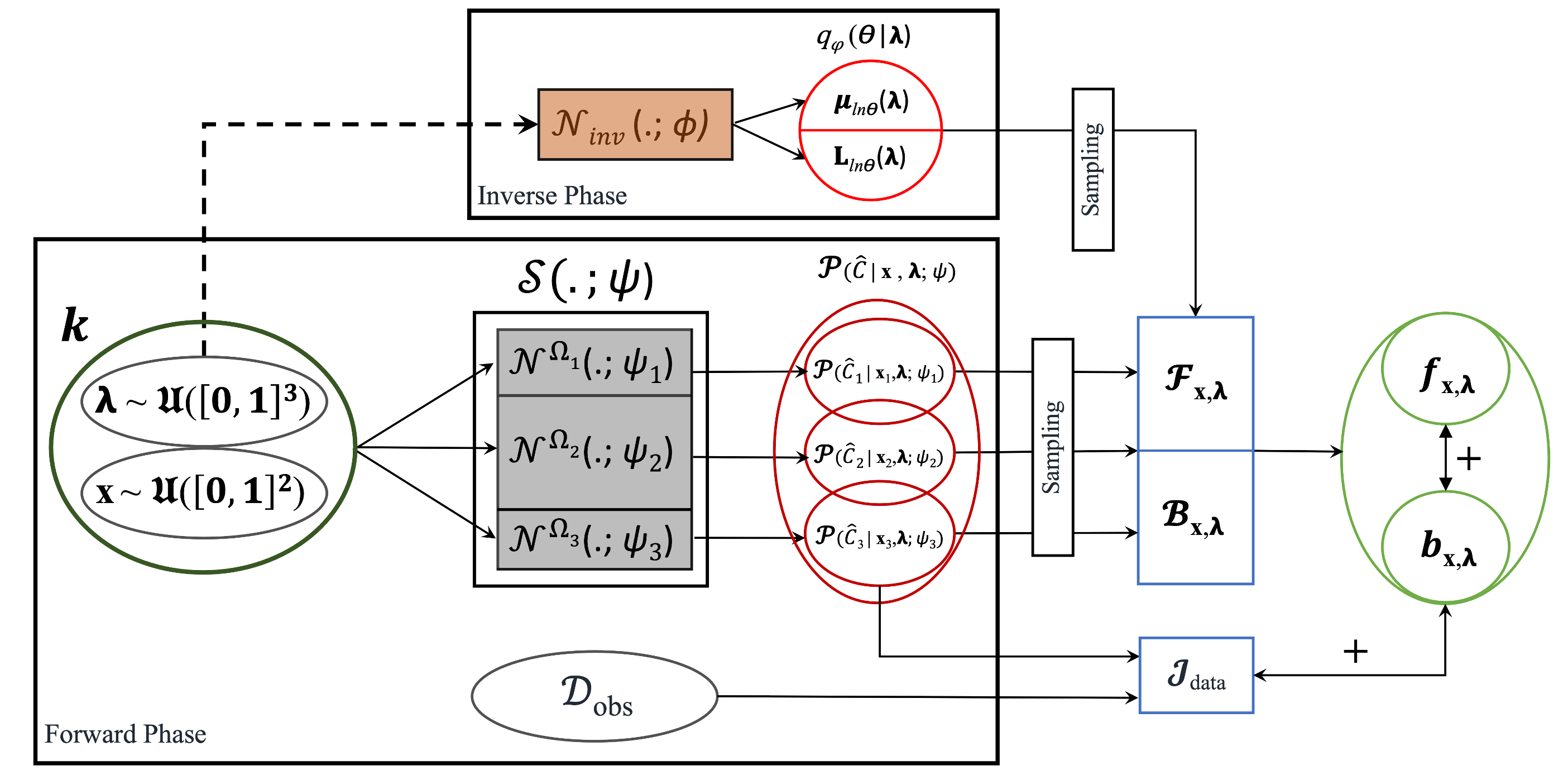}
    \caption{Schematic of the coupled CoDe-BPINN framework. \textbf{Inverse Phase:} The Parameter Network $\mathcal{N}_{inv}$ maps the operational design parameters $\boldsymbol{\lambda}$ onto the estimated statistics ($\boldsymbol{\mu}_{\ln\theta}, \mathbf{L}_{\ln\theta}$) of the physical parameters, $q_{\boldsymbol{\phi}}(\boldsymbol{\theta}|\boldsymbol{\lambda})$. \textbf{Forward Phase:} The Solver $\mathcal{S}$, composed of three Bayesian sub-networks ($\mathcal{N}^{\Omega_i}$), takes concatenated inputs ($\mathbf{x}, \boldsymbol{\lambda}$) and provides as outputs the concentration field, $\mathcal{P}(\hat{C} | \mathbf{x}, \boldsymbol{\lambda}; \boldsymbol{\psi})$. To form the composite loss, samples from both the solver and parameter network are fed into the PDE ($\mathcal{F}$) and Boundary ($\mathcal{B}$) operators to evaluate the physics-based residuals ($f_{\mathbf{x},\boldsymbol{\lambda}}, b_{\mathbf{x},\boldsymbol{\lambda}}$). At the same time, the solver's outputs at the outlet section of the system are compared against the sparse observational data ($\mathcal{D}_{obs}$) to compute a likelihood loss ($\mathcal{J}_{data}$). All components forming the loss are then aggregated into the final composite loss, $\mathcal{J}$ (detailed in Section~\ref{sec:ch4_composite_loss}), which is minimized during training.}
    \label{fig:ch4_architecture}
\end{figure}

\subsection{Physics Informed Neural Solver and Network Architecture}
\label{sec:ch4_architecture}
The proposed framework consists of two coupled components: the Forward Phase Solver $\mathcal{S}$ and the Inverse Phase Parameter Network $\mathcal{N}_{inv}$, as illustrated in Figure~\ref{fig:ch4_architecture}.

\subsubsection{Forward Phase: The Solver \texorpdfstring{($\mathcal{S}$)}{(S)}}
To capture the sharp gradients at the interfaces between the layers associated with the diverse porous materials forming the column, we employ an Conservative PINN (CPINN) architecture. The solver $\mathcal{S}$ is decomposed into the three independent sub-networks, $\{\mathcal{N}^{\Omega_i}(\cdot; \boldsymbol{\psi}_i)\}_{i=1}^3$, each corresponding to a given layer, $\{{\Omega_i}\}_{i=1}^3$, in the system. The main features of each of these sub-networks include the following:
\begin{itemize}
    \item \textbf{Architecture:} Each sub-network $\mathcal{N}^{\Omega_i}$ is a Bayesian Neural Network. The variational weights/biases $\boldsymbol{\psi}_i$ are modeled as independent Gaussian random variables $N(\mu_w, \sigma_w^2)$ ($N(\cdot,\cdot)$ denoting Gaussian).
    \item \textbf{Conditioning:} The sub-networks are conditioned via direct input concatenation. For a given point $\mathbf{x}$ in the domain and operational design vector $\boldsymbol{\lambda}$, the input is $\mathbf{k} = [\gamma(\mathbf{x}), \boldsymbol{\lambda}]$, where $\gamma(\cdot)$ is a Fourier feature embedding that is employed to mitigate spectral bias \citep{tancik2020fourier, ANAGNOSTOPOULOS2026107983}.
    \item \textbf{Output:} As outputs, each sub-network yields the parameters (mean and variance) of a normal distribution representing the model predictive uncertainty for the dimensionless concentration $\hat{C}$ within its corresponding subdomain $\Omega_i$:
    \begin{equation}
    \mathcal{P}(\hat{C} | \mathbf{x}, \boldsymbol{\lambda}; \boldsymbol{\psi}_i) = N\left( \mu_{\hat{C}}(\mathbf{x}, \boldsymbol{\lambda}; \boldsymbol{\psi}_i), \sigma_{\hat{C}}^2(\mathbf{x}, \boldsymbol{\lambda}; \boldsymbol{\psi}_i) \right)
    \end{equation}
\end{itemize}

\subsubsection{Inverse Phase: The Parameter Inference Network \texorpdfstring{($\mathcal{N}_{inv}$)}{(N-inv)}}
As stated in our Bayesian formulation (Section~\ref{sec:bayesian_formulation}), our goal is to approximate the posterior distribution of the physical parameters $\boldsymbol{\theta}$ with a variational distribution, $q_{\boldsymbol{\phi}}(\boldsymbol{\theta}|\boldsymbol{\lambda})$. This task is carried out by the \textbf{Parameter Inference Network}, $\mathcal{N}_{inv}$. The latter is a deterministic neural network, parameterized by weights $\boldsymbol{\phi}$, that maps the operational design parameters $\boldsymbol{\lambda}$ to the mean and covariance matrix of a multivariate Log-Normal distribution for $\boldsymbol{\theta}$. Relying on a Log-Normal distribution ensures that the inferred physical parameters remain strictly positive. The network outputs are:

\begin{equation}
[\boldsymbol{\mu}_{\ln\theta}(\boldsymbol{\lambda}; \boldsymbol{\phi}), \mathbf{L}_{\ln\theta}(\boldsymbol{\lambda}; \boldsymbol{\phi})] = \mathcal{N}_{inv}(\boldsymbol{\lambda};\boldsymbol{\phi})
\end{equation}
where $\boldsymbol{\mu}_{\ln\theta} \in \mathbb{R}^{d_\theta}$ is the vector whose entries are the means of the log-transformed parameters, and $\mathbf{L}_{\ln\theta} \in \mathbb{R}^{d_\theta \times d_\theta}$ is the lower-triangular Cholesky factor of the corresponding covariance matrix. This formulation enables the model to learn the correlation structure between parameters, which is critical to quantify practical non-identifiability. During training, samples of the physical parameter are generated using a reparameterization scheme, expressed as $\ln \boldsymbol{\theta} = \boldsymbol{\mu}_{\ln\theta} + \mathbf{L}_{\ln\theta} \mathbf{z}$ where $\mathbf{z} \sim N(0, I)$. This formulation enables end-to-end differentiability of the framework while allowing quantification of the parametric uncertainty in the inverse solution.

\subsubsection{Parameterization and Prior Beliefs}
\label{ch4_PriorB}
To regularize the inverse problem, we introduce an informative hyperprior, $p(\boldsymbol{\theta})$, which enters the KL divergence term of the ELBO (Eq.~\ref{eq:elbo}). This prior encodes our domain knowledge and constrains the (admissible) search space of the physical parameter vector $\boldsymbol{\theta}$, as defined in Section \ref{sec:ch4_problem_formulation}. We model $p(\boldsymbol{\theta})$ as a multivariate Log-Normal distribution, which is tantamount to assigning independent Gaussian priors to each log-parameter component $j=1, \dots, d_{\boldsymbol{\theta}}$: $\ln\theta_j \sim N(\mu_{p,j}, \sigma_{p,j}^2)$. Based on physical considerations and preliminary numerical simulations, these priors are defined as follows::

\begin{itemize}
   \item \textbf{Longitudinal Dispersivities ($a_{Ls}, a_{Lc}$):} Guided by the classically observed scaling that dispersivity is on the order of one-tenth of the characteristic travel length, priors for log-dispersivity are centered at $\mu_{p,1} = \ln(2.0)$ and $\mu_{p,2} = \ln(5.0)$ for the sand layers and for the more heterogeneous clay layer, respectively. The corresponding scales (standard deviations) are set to $\sigma_{p,j} = 0.5$ for $j=1,2$.

    \item \textbf{Sorption Capacity ($\beta$):} Preliminary batch (equilibrium) experiments suggest an adsorption capacity of $\beta \approx 60$ mg/g. Otherwise, effective capacity observed under dynamic column conditions is typically lower. This is mainly due to effects associated with, e.g., pore-scale flow heterogeneities (reflected through channeling) and kinetic limitations (associated with finite mass-transfer rates) that prevent the system from reaching equilibrium within the residence time of the flowing fluid. To account for these effects and provide sufficient flexibility in the inference process, we center the prior for $\ln\beta$ at a reduced value of $\mu_{p,4} = \ln(40)$ (where $\beta$ is expressed in mg/g). A relatively large standard deviation of $\sigma_{p,4} = 1.0$ is prescribed, enabling the Parameter Inference Network ($\mathcal{N}_{inv}$) to capture potential dependencies on operational conditions, such as flow rate, directly from the data.

    \item \textbf{Sorption Affinity ($\alpha$):} Preliminary batch equilibrium tests suggest an affinity coefficient $\alpha \approx 0.03$ L/mg. However, breakthrough curves obtained under dynamic column conditions are observed to exhibit significant spreading, indicative of a lower effective affinity. These discrepancies might arise from transport-related effects such as, e.g., dispersion, flow heterogeneity, and rate-limited sorption. To reflect these effects and maintain flexibility in the inference, we prescribe a weak prior for $\ln\alpha$ upon centering the latter at a substantially lower value of $\mu_{p,3} = \ln(3\times10^{-3})$ L/mg, with a comparatively large standard deviation of $\sigma_{p,3} = 1.0$. This broad prior enables the coupled framework to infer the appropriate balance between dispersive spreading and nonlinear adsorption (self-sharpening) effects that best explains the observed data.
\end{itemize}

\subsubsection{Physics-Informed Composite Loss} 
\label{sec:ch4_composite_loss}
Optimization of the parameters of all networks, $\Phi = \{\boldsymbol{\psi}, \boldsymbol{\phi}\}$, is performed through minimization of a composite loss function $\mathcal{J}$. The latter provides a tractable implementation of the negative ELBO derived in Section~\ref{sec:bayesian_formulation}. The negative log-likelihood contributions associated with the physical constraints are expressed as weighted Mean Squared Error (MSE) terms (as established in Section~\ref{sec:physics_likelihood}). This leads to the following scalar objective function:
\begin{equation}
\mathcal{J} = w_{data}\mathcal{J}_{data} + w_{pde}\mathcal{J}_{pde} + w_{bc}\mathcal{J}_{bc} + w_{ic}\mathcal{J}_{ic} + w_{int}\mathcal{J}_{int} + w_{kl}\mathcal{J}_{kl}
\label{eq:ch4_total_loss_detailed}
\end{equation}
Here, the boundary-related contribution from the theoretical formulation ($\mathcal{B}$) is decomposed into distinct terms corresponding to boundary ($\mathcal{J}_{bc}$), initial ($\mathcal{J}_{ic}$), and interface ($\mathcal{J}_{int}$) conditions. This decomposition allows for finer control of the optimization process through the weighting coefficients $w_{(\cdot)}$. The contributions appearing in Eq. \eqref{eq:ch4_total_loss_detailed} are defined in the following.

\paragraph{Residual Components.}
\begin{itemize}
    \item \textbf{Data Likelihood ($\mathcal{J}_{data}$):} This term represents the negative log-likelihood of the observational data, $-\ln p(\mathcal{D}_{obs} | \boldsymbol{\psi}, \boldsymbol{\theta})$. It is evaluated as the Negative Log-Likelihood (NLL) of the sparse breakthrough measurements at the column outlet, computed using the predictive distribution provided by the outlet sub-network $\mathcal{N}^{\Omega_3}$.

    \item \textbf{PDE Residuals ($\mathcal{J}_{pde}$):} This term corresponds to the negative log-likelihood associated with the PDE residuals, $-\ln p(\mathcal{F} | \cdot)$, as defined in Eq.~\ref{eq:physics_likelihood}. The expectation under the variational posterior is approximated via Monte Carlo sampling. At each training step, we draw samples of physical parameter values $\boldsymbol{\theta} \sim q_{\boldsymbol{\phi}}(\boldsymbol{\theta}|\boldsymbol{\lambda})$ and network weights $\boldsymbol{\psi} \sim q_{\boldsymbol{\psi}}(\boldsymbol{\psi})$, evaluate the residual $\mathcal{F}[\hat{C}]$, and minimize its magnitude. This contribution is computed independently across each of the subdomains of the column.

    \item \textbf{Boundary/Initial Conditions ($\mathcal{J}_{bc}, \mathcal{J}_{ic}$):} These terms represent the MSE counterparts of the log-likelihoods associated with the boundary and initial conditions, enforcing the constraints specified in Eq.~\ref{eq:ch4_bcs} on the corresponding sub-networks.

 \item \textbf{Interface Continuity ($\mathcal{J}_{int}$):} Within the domain-decomposed CPINN solver, continuity conditions are enforced at the material interfaces between adjacent layers. Notably, both the state ($\hat{C}$) and the total flux (denoted here as $J_{flux}$) required to satisfy continuity  across the material interfaces, to enforce mass conservation across internal boundaries. Considering vector ${\bf x}_{int} =[21, 71]$ mm to denote location of the interfaces along the column, the corresponding continuity constraint reads as:
    \begin{align}
    \mathcal{J}_{int} &= \sum_{k=1}^{2} \left( ||\hat{C}_{k}(x_{int,k}) - \hat{C}_{k+1}(x_{int,k})||^2 + ||J_{flux, k}(x_{int,k}) - J_{flux, k+1}(x_{int,k})||^2 \right)
    \end{align}
    where the total flux is given by $J_{flux} = v_x \hat{C} - D_L \partial_x \hat{C}$.

    \item \textbf{Regularization ($\mathcal{J}_{kl}$):} The regularization term, scaled by an annealing weigthing coefficient $w_{kl}$, consists of two distinct KL divergence contributions that penalize deviations of the learned variational distributions from the prior specifications introduced in Section~\ref{ch4_PriorB}.
        \begin{equation}
        \mathcal{J}_{kl} = \mathcal{J}_{kl, \mathcal{S}} + \mathcal{J}_{kl, \mathcal{N}_{inv}} =\sum_{i=1}^{3} D_{KL}[q_{\boldsymbol{\psi}_i}(\boldsymbol{\psi}_i) || p(\boldsymbol{\psi}_i)] + D_{KL}[q_{\boldsymbol{\phi}}(\boldsymbol{\theta}|\boldsymbol{\lambda}) || p(\boldsymbol{\theta})]
        \label{eq:kl_total}
        \end{equation}
    \begin{itemize}
        \item \textbf{Solver Weight Regularization:} For the BNN solver, we adopt a standard Gaussian prior over the weights, $p(\boldsymbol{\psi}) = N(0, I)$. The corresponding regularization term is given by the sum of the analytical KL divergences between the variational posterior $q_{\boldsymbol{\psi}_i}(\boldsymbol{\psi}_i)$ of each of the three sub-networks and its prior:
        \begin{equation}
            \mathcal{J}_{kl, \mathcal{S}} = \sum_{i=1}^3 D_{KL}[q_{\boldsymbol{\psi}_i}(\boldsymbol{\psi}_i) || p(\boldsymbol{\psi}_i)]
        \end{equation}

        \item \textbf{Physical Parameter Regularization:} we impose an informative prior, $p(\boldsymbol{\theta})$, modeled as a multivariate Log-Normal distribution as detailed in Section~\ref{ch4_PriorB}. The Parameter Inference Network ($\mathcal{N}_{inv}$) yields the key statistics of the variational distribution $q_{\boldsymbol{\phi}}(\ln\boldsymbol{\theta}|\boldsymbol{\lambda})$, which is also modeled as a multivariate Gaussian. The corresponding KL divergence between the variational posterior and the prior is evaluated in closed form as:
        \begin{equation}
        \begin{split}
            \mathcal{J}_{kl, \mathcal{N}_{inv}} = D_{KL}[q_{\boldsymbol{\phi}} || p] = \frac{1}{2} \bigg[ & \underbrace{\text{tr}(\boldsymbol{\Sigma}_p^{-1} \boldsymbol{\Sigma}_q)}_{\text{Trace Term}} + \underbrace{(\boldsymbol{\mu}_p - \boldsymbol{\mu}_q)^\top \boldsymbol{\Sigma}_p^{-1} (\boldsymbol{\mu}_p - \boldsymbol{\mu}_q)}_{\text{Mahalanobis Distance}} \\
            & - d_{\theta} + \underbrace{\ln \frac{|\boldsymbol{\Sigma}_p|}{|\boldsymbol{\Sigma}_q|}}_{\text{Log-Determinant Term}} \bigg]
        \end{split}
        \label{eq:multivariate_kl}
        \end{equation}
        where $\boldsymbol{\Sigma}_q = \mathbf{L}_{\ln\theta}\mathbf{L}_{\ln\theta}^\top$, and $d_{\theta}=4$ denotes the dimensionality of the physical parameter vector. This term is key in regularizing the inverse problem. The \textit{Trace Term} penalizes overly large posterior variances; the \textit{Mahalanobis Distance} anchors the mean estimate toward the prior (scaled by uncertainty); and the \textit{Log-Determinant Term} prevents the distribution from collapsing to a point estimate (corresponding to a degenerate solution).
    \end{itemize}
\end{itemize}

\paragraph{Probabilistic Interpretation of Loss Weights.}
We recall that in the probabilistic framework we consider, while observational data are naturally associated with measurement noise, the residuals of the governing equations (e.g., PDE, boundary, and initial conditions) are modeled as zero-mean Gaussian random variables (i.e., $p(\mathcal{F}) \propto \exp(-\frac{1}{2\sigma^2_{\mathcal{F}}} ||\mathcal{F}||^2)$) with constant variance. In this context, the associated variances do not represent measurement uncertainty. They are rather interpreted as tolerance parameters that control the extent at which each physical constraint is enforced. Under this assumption, maximizing the corresponding log-likelihoods is equivalent to minimizing weighted mean squared error (MSE) terms, where the corresponding weights are inversely proportional to the variance (or tolerance) $\sigma^2$ of each physical constraint, up to a constant factor:
\begin{equation}
w_{i} = \frac{1}{2\sigma_{i}^2} \quad \implies \quad \sigma_{i} = \frac{1}{\sqrt{2w_{i}}}
\label{eq:weight_variance_relation}
\end{equation}
Since all state variables are normalized to the interval $[0,1]$, the parameters $\sigma_i$ can be interpreted as relative tolerance levels for each constraint, expressed with respect to the characteristic scale of the (normalized) solution (i.e., relative to the inlet concentration $C_0$). Within this probabilistic framework, the corresponding weights therefore control the strength with which each contribution is enforced. In this study, we adopt a hierarchical weighting strategy to stabilize the inverse problem. Boundary, initial, and interface conditions are assigned the largest weights ($w_{bc}, w_{ic}, w_{int} = 8 \times 10^4$) corresponding to a stringent tolerance ($\sigma \approx 0.25\%$ of the unit interval), thereby ensuring strong enforcement of these constraints. The data is weighted as $w_{data} \approx 200$, reflecting a measurement uncertainty of $\approx 5\%$. Conversely, the PDE residual is assigned a moderately lower weight ($w_{pde} = 5 \times 10^3$), corresponding to a tolerance of $\approx 1\%$, which allows for controlled flexibility in satisfying the governing equations. Weighting of the KL regularization term, $w_{kl}$, is determined as described in SI Section 2. This soft-constrained formulation avoids an excessively stiff optimization landscape, thereby enabling the Parameter Inference Network to effectively explore the parameter space and identify dependencies that best reconcile physical consistency with observational data.

\subsection{Training via a Multi-Stage Curriculum}
\label{sec:ch4_training_strategy}
Directly optimizing the coupled forward-inverse problem is significantly challenging, primarily due to the \textit{cold start} issue. At this stage, the sub-networks composing $\mathcal{S}$ are not yet consistent across interfaces, and the Parameter Inference Network ($\mathcal{N}_{inv}$) produces essentially uninformative physical parameter estimates. Enforcing the full set of PDE constraints from the outset can therefore lead to optimization pathologies, such as the emergence of unphysical high-frequency artifacts that locally satisfy the differential operator while violating global boundary and data constraints.

To mitigate these issues, we employ a multi-stage curriculum learning strategy, extending the sequential training protocols introduced in our previous work on parameterized solvers \citep{panahi2026pids} to the present coupled inverse setting. Consistent with broader observations on PINN optimization challenges \citep{krishnapriyan2021characterizing, penwarden2023metalearning}, this approach stabilizes training by progressively increasing the complexity of the objective. In particular, it ensures that the network first learns to satisfy boundary and initial conditions before enforcing the full set of physics-based constraints. The procedure is formalized in Algorithm~\ref{alg:training_strategy_bpinn}.

\begin{algorithm}[!ht]
    \SetKwInput{KwData}{Input}
    \SetKwInput{KwResult}{Result}
    \SetKwComment{Comment}{/* }{ */}
    \caption{Three-Stage Curriculum Training for CoDe-BPINN}
    \label{alg:training_strategy_bpinn}
    
    \KwData{Operational design parameters $\boldsymbol{\lambda}$, Collocation points $\mathbf{x}$, Observational Data $\mathcal{D}_{obs}$}
    \KwResult{Optimized Solver weights $\boldsymbol{\psi}^*$, Parameter Inference Network weights $\boldsymbol{\phi}^*$}
    \BlankLine
    Initialize Solver weights $\boldsymbol{\psi}$ and Parameter Inference Network weights $\boldsymbol{\phi}$\;
    Freeze $\boldsymbol{\phi}$ ($\mathcal{N}_{inv}$) and fix physical parameters $\boldsymbol{\theta}$ to prior means\;
    
    \BlankLine
    \tcp*{\textbf{Stage 1: Boundary and Interface Initialization}}
    \For{$epoch \leftarrow 1$ \KwTo $E_1$}{
        Compute Boundary $\mathcal{J}_{bc}$ and Initial $\mathcal{J}_{ic}$\;
        Update $\boldsymbol{\psi}$ to minimize $\mathcal{J}_{Stage1} = w_{bc}\mathcal{J}_{bc} + w_{ic}\mathcal{J}_{ic}$\;
    }
    
    \BlankLine
    \tcp*{\textbf{Stage 2: Data Anchoring}}
    \For{$epoch \leftarrow 1$ \KwTo $E_2$}{
        Compute Data Likelihood $\mathcal{J}_{data}$ (NLL) at outlet\;
        Update $\boldsymbol{\psi}$ to minimize $\mathcal{J}_{Stage2} = \mathcal{J}_{Stage1} + w_{data}\mathcal{J}_{data}$\;
    }
    
    \BlankLine
    \tcp*{\textbf{Stage 3: Forward-Inverse Coupling}}
    Unfreeze $\boldsymbol{\phi}$ ($\mathcal{N}_{inv}$)\;
    \For{$epoch \leftarrow 1$ \KwTo $E_3$}{
        Update KL weight $w_{kl}$ via annealing schedule\;
        Sample physical parameters: $\boldsymbol{\theta} \sim q_{\boldsymbol{\phi}}(\boldsymbol{\theta} | \boldsymbol{\lambda})$\;
        Compute PDE Residuals $\mathcal{J}_{pde}$ and Interface loss $\mathcal{J}_{int}$ using sampled $\boldsymbol{\theta}$\;
        Compute Regularization $\mathcal{J}_{kl}$ for both $\boldsymbol{\psi}$ and $\boldsymbol{\phi}$\;
        Update $[\boldsymbol{\psi}, \boldsymbol{\phi}]$ to minimize $\mathcal{J}_{total} = \mathcal{J}_{Stage2} + w_{pde}\mathcal{J}_{pde} + w_{int}\mathcal{J}_{int} + w_{kl}\mathcal{J}_{kl}$\;
    }
    \KwRet{$\boldsymbol{\psi}, \boldsymbol{\phi}$}
\end{algorithm}

\paragraph{Stage 1: Boundary and Initial initialization.}
Training commences upon optimizing the solver $\mathcal{S}$ solely on boundary ($\mathcal{J}_{bc}$) and initial ($\mathcal{J}_{ic}$) conditions. The PDE residuals and interface continuity terms are deactivated during this phase. This enables the sub-networks to learn valid representations of the spatiotemporal boundaries for each subdomain.

\paragraph{Stage 2: Data Anchoring.}
Once the boundary representations are stable, we introduce the data likelihood term ($\mathcal{J}_{data}$). At this stage, the solver updates its posterior to fit the sparse observational data (i.e., concentration tempral history at the column outlet) while the PDE residual remains inactive. The model therefore operates as a probabilistic regressor, learning a solution manifold that is consistent with both boundary conditions and available observations. This stage effectively anchors the solution in data-informed regions, reducing risks of convergence to unphysical local minima when the physics-based constraints are subsequently enforced.

\paragraph{Stage 3: Forward-Inverse Coupling.}
In the final phase, the PDE residual ($\mathcal{J}_{pde}$) and interface continuity ($\mathcal{J}_{int}$) loss terms are activated, and the Parameter Inference Network ($\mathcal{N}_{inv}$) is incorporated into the training process, thereby closing the inverse loop. At this stage, the optimization must reconcile the data-anchored solution with the governing ADR equation. Since the solver already provides a good quality fit to the observations, minimizing the PDE residual primarily drives updates in the Parameter Inference Network, thus guiding the inference of the physical coefficients $\boldsymbol{\theta}$ required for physical consistency. Concurrently, the interface loss enforces continuity between sub-networks, ensuring global consistency of both the state (i.e., concentration) and the flux across the domain.

\subsection{Validation of Uncertainty Estimates}
\label{sec:ch4_uncertainty_quantification}

To assess model performance, we employ a comparative Monte Carlo analysis using both the trained CoDe-BPINN and a reference FDM solver (see Supplementary Information, Section 1). The total predictive uncertainty of the CoDe-BPINN is estimated by generating an ensemble of $T$ stochastic forward evaluations. For each realization $t$, we sample both the network weights $\boldsymbol{\psi}_t \sim q_{\boldsymbol{\psi}}(\boldsymbol{\psi})$ and the physical parameters $\boldsymbol{\theta}_t \sim q_{\boldsymbol{\phi}}(\boldsymbol{\theta}|\boldsymbol{\lambda})$. The resulting ensemble of results, $\{\hat{C}(\mathbf{x}; \boldsymbol{\psi}_t, \boldsymbol{\theta}_t)\}_{t=1}^T$, captures the total predictive uncertainty, encompassing variability due to both parameter inference and the neural surrogate approximation. To establish a reference ground truth for the parametric contribution, we rely on the trained Parameter Inference Network ($\mathcal{N}_{inv}$) to generate input samples for the FDM simulator. We draw realizations of physical parameters $\boldsymbol{\theta}_t \sim q_{\boldsymbol{\phi}}(\boldsymbol{\theta}|\boldsymbol{\lambda})$ and evaluate the FDM solver for each sample. The variance of the ensuing (FDM-based) ensemble provides an estimate of the uncertainty propagated solely from the inferred physical parameters (hereafter denoted as $\sigma_{\theta}^2$).

Comparing these two ensembles provides a basis for validation of the CoDe-BPINN results. In principle, the total (predictive) uncertainty of the CoDe-BPINN should encompass the parametric uncertainty obtained from the FDM-based ensemble, as it additionally accounts for the approximation error of the neural surrogate. In regions with limited observational data, uncertainty is expected to be dominated by the parametric component, reflecting the inherent non-uniqueness of the inverse problem. Conversely, near the data-rich outlet, the parametric uncertainty should reduce significantly, leaving only the residual measurement noise and surrogate-related uncertainty. This comparison offers a robust assessment of the reliability of the proposed modeling framework.

\section{Results and Discussion}
\label{sec:ch4_results}

In this section, we evaluate the performance of the proposed CoDe-BPINN framework. The analysis is structured to assess two distinct capabilities: (1) the accuracy of the \textbf{Forward Surrogate} (the solver $\mathcal{S}$) in reproducing the spatiotemporal concentration field compared to a high-fidelity numerical reference, and (2) the ability of the \textbf{Inverse Parameter Network} ($\mathcal{N}_{inv}$) to estimate the unknown physical parameters and their  dependencies on the operational design parameters.

\subsection{Experimental Dataset and Operational design Space}
\label{sec:ch4_dataset}
The CoDe-BPINN framework was trained on the dataset $\mathcal{D}_{obs}$ comprising 9 experimental breakthrough curves (BTCs) obtained from the experimental setup described in Section~\ref{sec:experimental_setup}. 

\paragraph{Data Preprocessing and Normalization.}
As illustrated in the network architecture (Fig.~\ref{fig:ch4_architecture}), all network inputs are normalized to ensure numerical stability during training. Spatial coordinates $x$ are scaled by the column length ($L=92$ mm), time $t$ is normalized by the maximum experimental duration ($T_{max}$), and the operational design parameter vector $\boldsymbol{\lambda} = [\varepsilon_{\Omega_2}, q, C_0]^\top$ is mapped to the support $[0, 1]$ via min-max normalization. Importantly, while the networks operate in this normalized space, the physics-informed loss terms (i.e., the PDE and Interface Continuity residuals) are evaluated in the actual \textit{physical domain}. To ensure consistency, network outputs are transformed back to physical units during training, and the gradients obtained through automated differentiation are appropriately rescaled (e.g., $\frac{1}{L} \frac{\partial}{\partial x_{\mathrm{norm}}}$) so that the governing equations (Eq.~\ref{eq:ch4_strong_form}) are enforced in their correct physical form.

\paragraph{Fixed System Parameters.}
While the operational design vector $\boldsymbol{\lambda}$ defines the diverse experimental conditions, the geometric configuration (Section \ref{exp_geometric}) and material properties of the non-reactive components are kept constant across all experiments. The inert sand layers ($\Omega_1$ and $\Omega_3$) are characterized by a fixed porosity of $\varepsilon_{\Omega_1} = \varepsilon_{\Omega_3} = 0.43$, bulk density of the porous medium being set to $\rho_b = 1.5 \times 10^{-3}$ g/mm$^3$). These quantities are treated as known constants and are embedded directly within the PDE residual operator during training.

\subsection{Surrogate Modeling Performance}
\label{sec:ch4_surrogate_performance}
\begin{figure}[!ht]
    \centering
    \includegraphics[width=\textwidth]{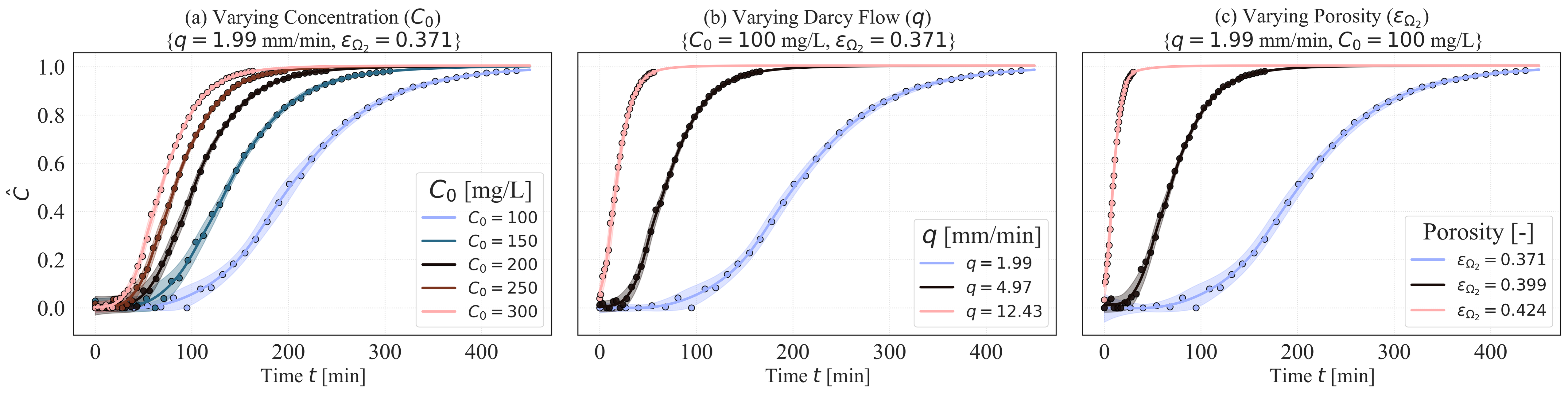}
    \caption{Comparison of CoDe-BPINN surrogate results (solid curves) with available experimental observations (dots) at the column outlet ($x=L$). Concentrations are expressed in dimensionless form as $\hat{C} = C/C_0$. Shaded regions denote the total predictive uncertainty (corresponding to 95\% CI) estimated via Monte Carlo sampling. The model accurately captures the non-linear variations in breakthrough time and breakthrough curve shape associated with changes in the operational variables, i.e., (a) inlet concentration $C_0$, (b) Darcy flux $q$, and (c) porosity $\varepsilon_{\Omega_2}$.}
    \label{fig:ch4_btc_panorama}
\end{figure}

The primary objective of the Forward Phase is to learn a continuous, parameterized solution manifold $\hat{C}(\mathbf{x}, \boldsymbol{\lambda})$ that accurately reproduces the observed transport dynamics. Figure~\ref{fig:ch4_btc_panorama} depicts a comprehensive comparison of the CoDe-BPINN results against the experimental data across the three axes of variation, illustrating the ability of the model to capture the observed breakthrough behavior.

These results demonstrate that the solver $\mathcal{S}$ has successfully learned to characterize the complex, non-linear dependencies governing the system.
As shown in Fig.~\ref{fig:ch4_btc_panorama}a, the model accurately captures the concentration-dependent behavior, including the characteristic self-sharpening of the breakthrough front and the earlier breakthrough times observed at higher inlet concentrations ($C_0$). As $C_0$ increases from 100 to 300 mg/L, adsorption sites saturate more rapidly, leading to a leftward shift of the breakthrough curve. The CoDe-BPINN reproduces these sharp fronts without introducing spurious oscillations, highlighting the effectiveness of the Fourier embedding in representing high-frequency features.
In Fig.~\ref{fig:ch4_btc_panorama}b, variations in Darcy flux ($q$) induce substantial changes in residence time. The model successfully captures the transition from advection-dominated transport (steep breakthrough curve at $q=12.43$ [mm/min]) to dispersion-dominated regimes (broader curves at $q=1.99$ [mm/min]). Dynamics resulting from our modeling framework  remain consistent with experimental observations across timescales spanning approximately one order of magnitude.
Similarly, Fig.~\ref{fig:ch4_btc_panorama}c illustrates the impact of the porosity of the middle layer, $\varepsilon_{\Omega_2}$, which serves as a proxy for the adsorbent mass fraction. The model correctly learns that lower porosity (corresponding to higher clay content, $\varepsilon_{\Omega_2}=0.371$) leads to significantly increased retardation capacity compared to a sand-dominated matrix ($\varepsilon_{\Omega_2}=0.424$), shifting breakthrough time from $\approx 30$ minutes to $ \approx 440$ minutes.

Furthermore, the uncertainty bands (shaded regions) tightly envelop experimental observations, indicating that the Bayesian solver captures the underlying structure.


\subsection{Surrogate Accuracy and Uncertainty Calibration}
\label{sec:ch4_surrogate_accuracy}

A primary objective of our framework is to develop a CoDe-BPINN that enables unknown parameter inference while at the same time serving as an accurate and reliable forward surrogate. To assess this capability, we compare the predictive distribution obtained through CoDe-BPINN against a high-fidelity Monte Carlo ensemble generated through a Finite Difference Method (FDM) solver (as detailed in SI, Section 1). The FDM simulations are driven by realizations of the physical parameters sampled from the posterior distribution learned by the Parameter Network ($\mathcal{N}_{inv}$), thus enabling a consistent evaluation of both the predictive mean and the propagated uncertainty.

\begin{figure}[!ht]
    \centering
    \includegraphics[width=\textwidth]{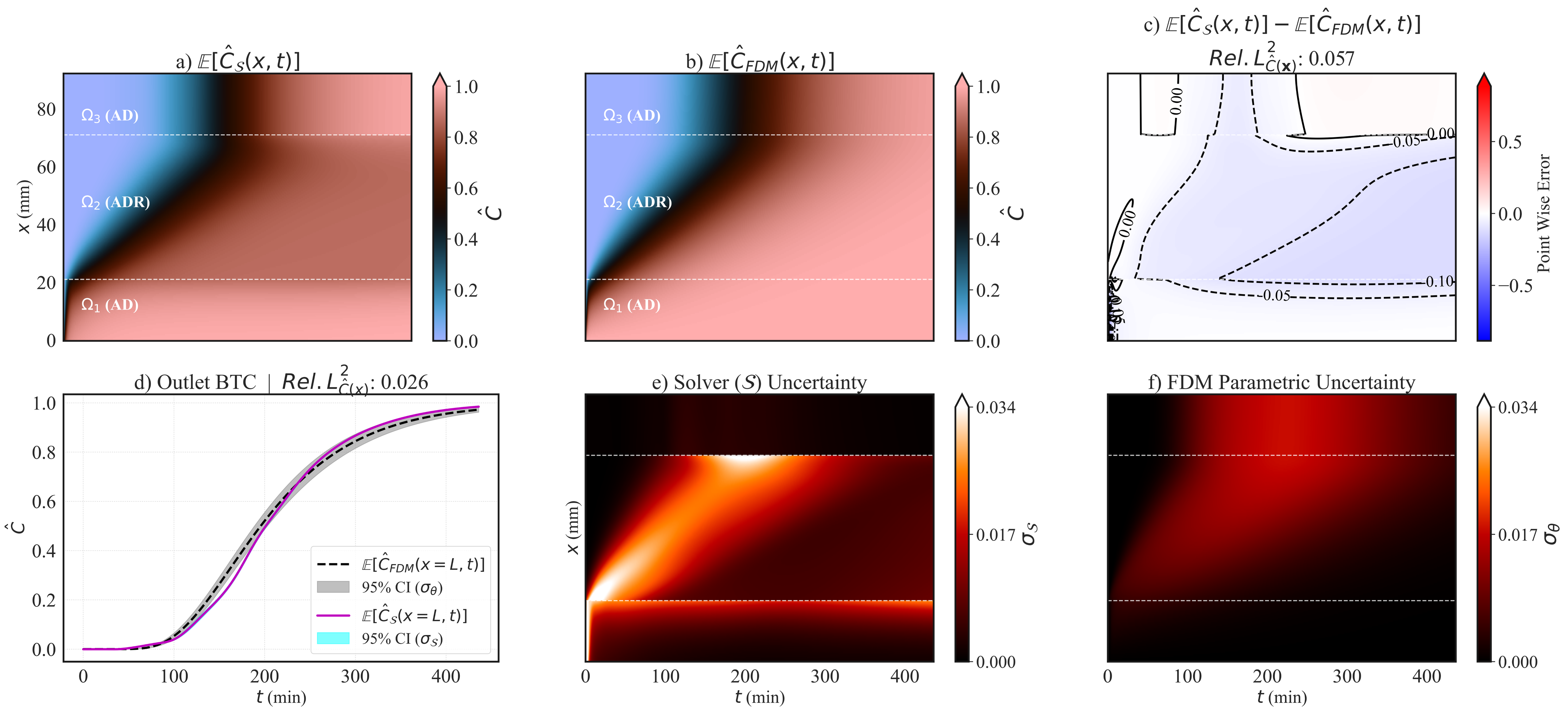}
    \caption{\textbf{Ensemble-based validation of the CoDe-BPINN surrogate.} Comparison between CoDe-BPINN results against an FDM-based Monte Carlo ensemble ($N=1000$) for the baseline setting $\boldsymbol{\lambda} = [0.371, 1.99, 100]$. \textbf{(a)} Mean CoDe-BPINN field $\mathbb{E} [\hat{C}_{\mathcal{S}} (x,t)]$. \textbf{(b)} Reference mean field from the FDM-based ensemble $\mathbb{E} [\hat{C}_{FDM} (x,t)]$. \textbf{(c)} Point-wise difference between the two mean fields. \textbf{(d)} Comparison of the outlet breakthrough curves and related uncertainty. \textbf(e) Spatiotemporal distribution of the CoDe-BPINN epistemic uncertainty ($\sigma_{\mathcal{S}}$). \textbf(f) Spatiotemporal distribution of the parameter-induced uncertainty obtained from the FDM ensemble ($\sigma_{\boldsymbol{\theta}}$).}
    
    \label{fig:ch4_verification_ensemble}
\end{figure}

Figure~\ref{fig:ch4_verification_ensemble} illustrates this comparison for the baseline experiment test case ($\boldsymbol{\lambda} = [\varepsilon_{\Omega_2}=0.371, q=1.99 \text{ mm/min}, C_0=100 \text{ mg/L}]$).

\begin{figure}[!ht]
    \centering
    \includegraphics[width=\textwidth]{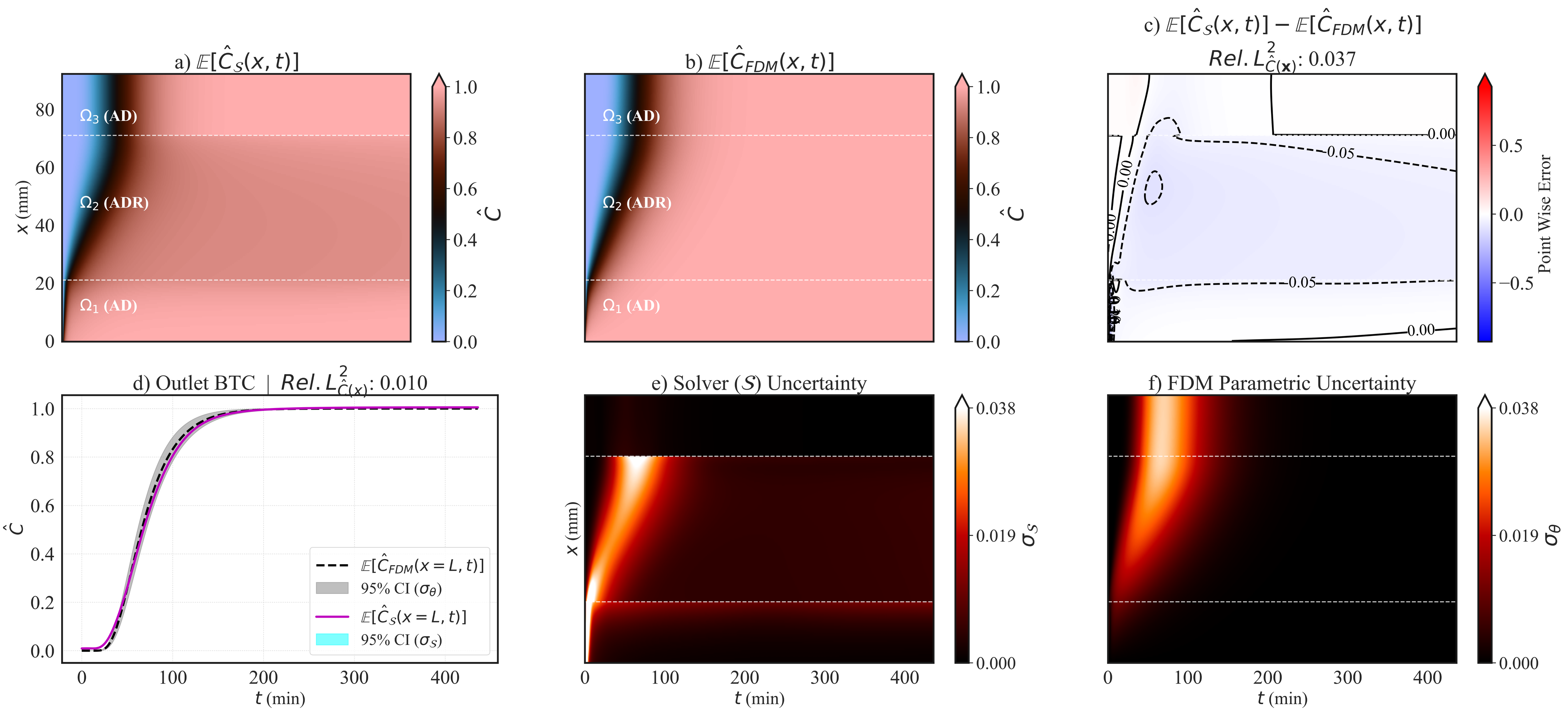}
    \caption{\textbf{Ensemble-based validation for an intermediate clay content of the adsorbent layer ($\varepsilon_{\Omega_2}=0.399$).} Comparison between CoDe-BPINN results and an FDM ensemble ($N=1000$) for the experiment ID 9 (see Table \ref{tab:exp_matrix}), corresponding to a medium clay content configuration ($\boldsymbol{\lambda} = [0.399, 1.99, 100]$. \textbf{(a)} Mean CoDe-BPINN field $\mathbb{E} [\hat{C}_{\mathcal{S}} (x,t)]$. \textbf{(b)} Reference mean field from the FDM ensemble $\mathbb{E} [\hat{C}_{FDM} (x,t)]$. \textbf{(c)} Point-wise difference between the mean fields. \textbf{(d)} Comparison of the outlet breakthrough curves and related uncertainty. \textbf(e) Spatiotemporal distribution of the CoDe-BPINN estimated uncertainty ($\sigma_{\mathcal{S}}$). \textbf(f) Spatiotemporal distribution of the parameter-induced uncertainty from the FDM ensemble ($\sigma_{\boldsymbol{\theta}}$).}
    \label{fig:ch4_verification_ensemble_poro2}
\end{figure}

\paragraph{Accuracy of Mean Predictions.}
The spatiotemporal mean field inferred by our CoDe-BPINN (Fig.~\ref{fig:ch4_verification_ensemble}a) exhibits excellent visual and quantitative agreement with the corresponding FDM ensemble mean (Fig.~\ref{fig:ch4_verification_ensemble}b). Both approaches accurately reproduce the key physical mechanisms, including rapid advective transport in the inert sand layers ($\Omega_1, \Omega_3$) and significant retardation within the reactive clay layer ($\Omega_2$). The point-wise difference map (Fig.~\ref{fig:ch4_verification_ensemble}c) reveals that the largest discrepancies are localized along the sharp concentration front, where steep gradients pose challenges for neural approximators. Nevertheless, the overall relative $Rel. L^2_{\hat{C}} (\mathbf{x})$ between the two mean fields is only 4.3\%, confirming that the CoDe-BPINN has recovered a physically consistent solution manifold. 
This agreement is further supported at the outlet (Fig.~\ref{fig:ch4_verification_ensemble}d), where the CoDe-BPINN mean breakthrough curve (magenta) closely matches its FDM-based reference counterpart (black dashed), with $Rel. L^2_{\hat{C}} (x=L,t) = 1.9\%$. The same panel also illustrates the total predictive uncertainty of the solver $\mathcal{S}$ (95\% CI, $\sigma_{total}^\mathcal{S}$, shown in shaded cyan) alongside the reference \textit{parametric uncertainty} propagated through the FDM solver (95\% CI, $\sigma_{\theta}$; shown in shaded grey). As detailed in the uncertainty quantification procedure (Section~\ref{sec:ch4_uncertainty_quantification}), $\sigma_{\theta}$ is obtained upon evaluating the deterministic FDM solver over realizations of $\boldsymbol{\theta}$ sampled from the learned variational distribution $q_{\boldsymbol{\phi}}(\boldsymbol{\theta}|\boldsymbol{\lambda})$.

\paragraph{Assessment and Decomposition of Predictive Uncertainty }
Our Bayesian framework allows quantifying and decompose predictive uncertainty. Figure~\ref{fig:ch4_verification_ensemble}e depicts predictive uncertainty stemming from the CoDe-BPINN (denoted as $\sigma_{\mathcal{S}}$ in the figure), which reflects the combined effect of two sources, respectively corresponding to $(i)$ \textit{surrogate uncertainty} (arising from variability in the solver weights) and $(ii)$ \textit{parametric uncertainty} (associated with the inferred physical parameters). In contrast, Figure~\ref{fig:ch4_verification_ensemble}f depicts the \textbf{parametric uncertainty} $\sigma_{\theta}$ obtained upon propagating the same parameter distribution through the reference FDM solver. For consistency, the FDM ensemble is driven by samples drawn from the posterior $q_{\boldsymbol{\phi}}(\boldsymbol{\theta}|\boldsymbol{\lambda})$ inferred by the Parameter Inference Network ($\mathcal{N}_{inv}$).

A key observation is the strong structural agreement between these two uncertainty fields. Both exhibit localized amplification along the reactive front within the clay layer ($\Omega_2$), where the solution is most sensitive to the (uncertain) sorption parameters $(\alpha, \beta)$. At the system outlet (Fig.~\ref{fig:ch4_verification_ensemble}d), the total uncertainty estimated by the CoDe-BPINN (magenta band) consistently envelops the reference parametric uncertainty obtained from the FDM ensemble (grey band). The slightly broader spread of the CoDe-BPINN reflects the additional contribution of surrogate approximation uncertainty . The close correspondence between these uncertainty estimates in the data-constrained (boundary) region provides strong evidence that the inferred parameter distribution accurately captures the variability required to reproduce the observed system behavior, thereby yielding a strong validation basis for the overall probabilistic inversion framework.

\paragraph{Generalization Across the Operational Design Space of Porosity.}
To further assess the generalization capability of the framework, we perform the same  analysis for a scenario corresponding to an intermediate clay fraction ($\varepsilon_{\Omega_2} = 0.399$). This latter scenario lies between previously observed configurations in the porosity dimension of the operational design space. The ensuing results are depicted in Figure~\ref{fig:ch4_verification_ensemble_poro2}.

Our CoDe-BPINN is seen to exhibit strong agreement with the reference solution also in this case, with a relative $L^2$ error of 2.8\% between its inferred mean field and the FDM ensemble mean. This accuracy is further confirmed at the outlet (Fig.~\ref{fig:ch4_verification_ensemble_poro2}d), where the CoDe-BPINN mean breakthrough curve closely matches its FDM counterpart (yielding $Rel. L^2_{\hat{C}} (x=1,t)$ = 0.8\%). The model correctly captures the shift in transport dynamics associated with this reduced clay content. The ensuing reduced sorption capacity leads to earlier breakthrough, this behavior being consistently reflected in both the spatiotemporal concentration fields and the breakthrough curve at the outlet. The corresponding uncertainty fields (comparing CoDe-BPINN predictive uncertainty, $\sigma_{\mathcal{S}}$, and the parameter-driven uncertainty, $\sigma_{\boldsymbol{{\theta}}}$ obtained from the FDM ensemble) again display strong agreement, with uncertainty localized along the reactive front. This consistent performance across different regions of the operational design space demonstrates both the robustness of the learned surrogate and the reliability of the inferred parameter distributions.

\subsection{Analysis of Learned Parameter Dependencies}
\label{sec:ch4_constitutive_discovery}
\begin{figure}[!ht]
    \centering
    \includegraphics[width=\textwidth]{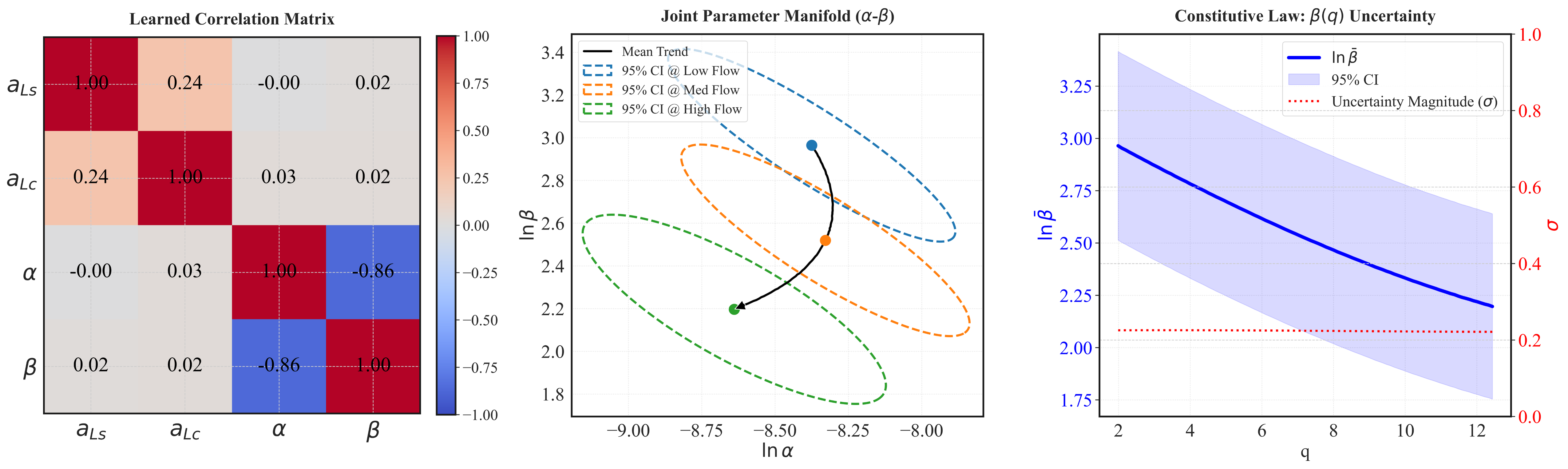}
    \caption{Multivariate analysis of the inferred physical parameters. \textbf{(Left)} The learned correlation matrix $\mathbf{R}_{\ln\theta}$ reveals a largely decoupled structure between dispersive ($a_{Ls}, a_{Lc}$) and sorption ($\alpha [L/mg], \beta [mg/g]$) parameters, together with a pronounced negative correlation ($\rho=-0.54$) between sorption affinity $\alpha$ and capacity $\beta$. \textbf{(Middle)} Evolution of the joint distribution in the $\ln\alpha$-$\ln\beta$ plane as Darcy flux $q$ increases from 1.99 to 12.43 mm/min. Dashed ellipses denote the corresponding 95\% confidence regions, illustrating the geometry of the inferred parameter uncertainty. \textbf{(Right)} Learned functional relationship, $\beta(q)$, showing a monotonic reduction of effective sorption capacity with increasing flow rate, together with uncertainty bounds.}
    \label{fig:ch4_multivariate_analysis}
\end{figure}

Beyond accurate state estimation, a key objective of this work is to uncover the underlying physical dependencies governing the reactive transport process. This is achieved by analyzing the outputs of the Multivariate Parameter Network ($\mathcal{N}_{inv}$), which provides a structured mapping between the physical parameters ($\boldsymbol{\theta}$) and the operating conditions ($\boldsymbol{\lambda}$). Through this analysis, it is possible to disentangle the coupled dependencies between transport and sorption properties and the experimental controls. Figure~\ref{fig:ch4_multivariate_analysis} provides a detailed characterization of the learned parameter manifold for the baseline experimental condition (High Clay, $\varepsilon_{\Omega_2}=0.371$; Low Concentration, $C_0=100$ mg/L).

\begin{figure}[!ht]
    \centering
    \includegraphics[width=\textwidth]{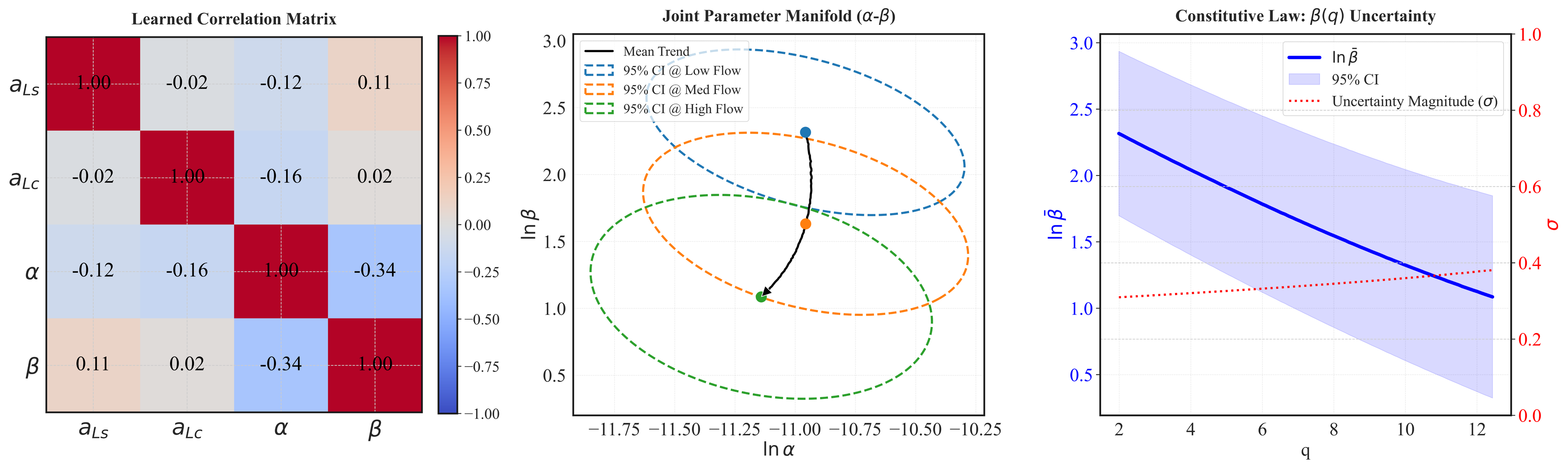}
    \caption{Extrapolation to an Unseen Operating Scenario. Multivariate analysis for a configuration characterized by minimum clay content ($\varepsilon_{\Omega_2}=0.424$) and maximum inlet concentration ($C_0=300$ mg/L). The model yields physically consistent trends, indicating a marked reduction in the overall effective sorption strength ($\alpha\beta$) driven by a decrease in both affinity and capacity. The progressive widening of the uncertainty bands at higher flow rates reflects increased uncertainty, consistent with extrapolation beyond the training domain.}
    \label{fig:ch4_multivariate_analysis_extrap}
\end{figure}

\begin{table}[!ht]
    \centering
    \caption{Inferred effective parameters for the nine experimental cases. Values are reported as mean $\pm$ standard deviation of the posterior distribution predicted by the Parameter Inference Network. The operational variables are Darcy flux ($q$), inlet concentration ($C_0$), and middle-layer porosity ($\varepsilon_{\Omega_2}$). The inferred parameters are sand and clay dispersivities ($a_{Ls}$ and $a_{Lc}$), sorption affinity ($\alpha$), and sorption capacity ($\beta$).}
    \label{tab:inferred_params}
    \resizebox{\textwidth}{!}{
    \begin{tabular}{c c c c | c c c c}
    \toprule
    \multicolumn{4}{c}{\textbf{Operational Parameters}} &
    \multicolumn{4}{c}{\textbf{Inferred Physical Parameters (Mean $\pm$ Std. Dev.)}} \\
    \cmidrule(lr){1-4} \cmidrule(lr){5-8}
    \textbf{Exp.} &
    $\boldsymbol{q}$ &
    $\boldsymbol{C_0}$ &
    $\boldsymbol{\varepsilon_{\Omega_2}}$ &
    $\boldsymbol{a_{Ls}}$ &
    $\boldsymbol{a_{Lc}}$ &
    $\boldsymbol{\alpha}$ &
    $\boldsymbol{\beta}$ \\
    \textbf{ID} &
    [mm/min] &
    [mg/L] &
    [-] &
    [mm] &
    [mm] &
    [L/mg] &
    [mg/g] \\
    \midrule
    1 & 1.989 & 100 & 0.371 &
    $1.710 \pm 0.180$ &
    $3.488 \pm 0.525$ &
    $(2.380 \pm 0.585)\times10^{-4}$ &
    $19.833 \pm 4.529$ \\

    2 & 4.974 & 100 & 0.371 &
    $2.847 \pm 0.253$ &
    $4.762 \pm 0.629$ &
    $(2.550 \pm 0.636)\times10^{-4}$ &
    $15.311 \pm 3.519$ \\

    3 & 12.434 & 100 & 0.371 &
    $10.362 \pm 0.599$ &
    $12.752 \pm 1.150$ &
    $(1.830 \pm 0.491)\times10^{-4}$ &
    $9.230 \pm 2.105$ \\

    4 & 1.989 & 150 & 0.371 &
    $1.725 \pm 0.193$ &
    $3.764 \pm 0.605$ &
    $(1.790 \pm 0.456)\times10^{-4}$ &
    $18.035 \pm 3.986$ \\

    5 & 1.989 & 200 & 0.371 &
    $1.705 \pm 0.209$ &
    $4.024 \pm 0.708$ &
    $(1.440 \pm 0.373)\times10^{-4}$ &
    $16.879 \pm 3.746$ \\

    6 & 1.989 & 250 & 0.371 &
    $1.671 \pm 0.225$ &
    $4.275 \pm 0.808$ &
    $(1.200 \pm 0.327)\times10^{-4}$ &
    $16.040 \pm 3.682$ \\

    7 & 1.989 & 300 & 0.371 &
    $1.627 \pm 0.240$ &
    $4.535 \pm 0.943$ &
    $(1.020 \pm 0.295)\times10^{-4}$ &
    $15.533 \pm 3.658$ \\

    8 & 1.989 & 100 & 0.424 &
    $0.997 \pm 0.267$ &
    $5.516 \pm 1.451$ &
    $(0.430 \pm 0.132)\times10^{-4}$ &
    $12.346 \pm 2.932$ \\

    9 & 1.989 & 100 & 0.399 &
    $1.302 \pm 0.225$ &
    $4.351 \pm 0.868$ &
    $(0.980 \pm 0.271)\times10^{-4}$ &
    $15.448 \pm 3.464$ \\
    \bottomrule
    \end{tabular}}
\end{table}

\paragraph{Structure of Learned Parameter Correlations.}
The learned correlation matrix (Fig.~\ref{fig:ch4_multivariate_analysis}, Left) provides quantitative evidence that the CoDe-BPINN has effectively partitioned the governing physics. The matrix exhibits a distinct block-diagonal structure, indicating that the model has identified two weakly coupled groups of parameters, i.e., dispersive ($a_{Ls}, a_{Lc}$) and reactive ($\alpha, \beta$) parameters, each governing distinct aspects of the transport dynamics. Within the sorption subspace, a pronounced negative correlation ($\rho_{\alpha\beta} = -0.86$) emerges between the Langmuir affinity $\alpha$ and capacity $\beta$. This result reflects the observation that for a given breakthrough behavior, an increase in affinity must be offset by a decrease in capacity, and vice versa. The CoDe-BPINN captures this trade-off through a negative off-diagonal structure in the learned Cholesky factor $\mathbf{L}_{\ln\theta}$. Conversely, the dispersive parameters exhibit only  a weak positive correlation ($\rho_{a_{Ls}a_{Lc}} = 0.24$) between the sand and clay dispersivities. This suggests that the data constrain the dispersivity in each layer separately, with only weak interaction between them. The corresponding inferred mean values of the physical parameters ($\boldsymbol{\mu}_{\theta}$) across the nine experimental configurations are listed in Table~\ref{tab:inferred_params}.

\paragraph{Analysis of the Inferred Parameter Manifold.}
The joint evolution of the inferred mean parameters (Fig.~\ref{fig:ch4_multivariate_analysis}, Middle) reveals a non-trivial dynamic response to variations in hydrodynamic conditions. As Darcy flux $q$ increases (progressing from the blue to the green ellipse in Figure \ref{fig:ch4_multivariate_analysis}), the system transitions from a regime characterized by higher sorption capacity ($\ln \beta \approx 3.0$) to one with reduced capacity ($\ln \beta \approx 2.2$). The behavior of sorption affinity, $\alpha$, is notably non-monotonic. At low to intermediate flow rates, $\alpha$ increases slightly, suggesting enhanced mass transfer to the adsorbent surface, whereas reduced residence time becomes dominant at higher flow rates, leading to a marked decrease in both effective affinity ($\alpha$) and capacity ($\beta$). This resulting \textit{hook-shaped} trajectory indicates that the framework captures a non-linear trade-off between transport and sorption processes, rather than a simple monotonic relationship. The orientation of the uncertainty ellipses (that are aligned along the negative diagonal), reflects the strong anti-correlation between $\alpha$ and $\beta$. This pattern indicates that the model explores a direction of parameter equifinality, adjusting the balance between affinity and capacity to preserve consistency with the observed transport dynamics as flow conditions vary.

\paragraph{Learned Functional Dependence Between Flow Rate and Sorption Capacity}
To isolate the relationship between flow conditions and sorption capacity, we examine the marginal distribution obtained considering the results obtained in plane $q$ - $\beta$ (Fig.~\ref{fig:ch4_multivariate_analysis}, Right), from which a functional dependence  $\beta(q)$ can be inferred. The model identifies a clear monotonic decrease in effective sorption capacity with increasing Darcy flux, the effective median value of $\beta$ decreasing from $\approx 20$ mg/g to $\approx 9$ mg/g. This trend is consistent with a conceptual picture according to which reduced residence times at higher flow velocities tend to limit intra-particle diffusion, thereby restricting access to adsorption sites within microporous structures and effectively lowering the observable capacity. The associated uncertainty (red dotted curve in Figure \ref{fig:ch4_multivariate_analysis}) remains consistently low and stable across the entire range of flow rates, suggesting that this relationship is robustly supported by the data. It is also important to note that this behavior may partially reflect limitations of the adopted equilibrium sorption model, which does not explicitly account for kinetic or mass transfer effects. Incorporating such processes could provide a more accurate representation of the system, particularly at higher flow rates.

\paragraph{Extrapolation to Untrained Regimes.}
To assess generalization limits of the inferred model, we consider a scenario outside the training domani within the operational design space and corresponding to minimum clay content in the intermediate layer (i.e., maximum porosity, $\varepsilon_{\Omega_2}=0.424$) and maximum inlet concentration ($C_0=300$ mg/L). This configuration represents a joint extrapolation in the operational design space, as the model was not trained on data combining these conditions. The ensuing results are depicted in Figure~\ref{fig:ch4_multivariate_analysis_extrap}. 

Relative to the baseline case, the model projects a significant reduction in overall sorption strength, driven by decreases in both effective affinity ($\alpha$) and capacity ($\beta$). For instance, at a low flow rate ($q \approx 2$ mm/min), the inferred effective sorption coefficient ($\alpha\beta$) decreases from $\approx 0.0067$ in the high-clay baseline configuration to $\approx 0.00016$ in the low-clay scenario, corresponding to a reduction exceeding 95\%. While experimental data are not available for direct validation in this specific regime, this trend is consistent from a physical standpoint, as reduced adsorbent mass (associated with higher porosity) naturally lowers the total sorption capacity ($\beta$), while altered flow conditions likely reduce contact efficiency, leading to a reduced effective affinity ($\alpha$).

Furthermore, the model retains the monotonic decay of $\beta$ with increasing flow rate. Importantly, the \textit{total predictive uncertainty} (as reflected through the 95\% CI) expands noticeably at higher flow rates relative to the training regime. This suggests that the CoDe-BPINN appropriately identifies this region as weakly constrained by information. This behavior, which combines physically consistent extrapolations with increased uncertainty, suggests that the framework remains robust and avoids overconfident predictions in unobserved regions of the operational design spaces.

\subsection{Learned Parameter Manifolds in the Operational Design Space}
\label{sec:ch4_constitutive_manifolds}

\begin{figure}[!ht]
    \centering
    \includegraphics[width=\textwidth]{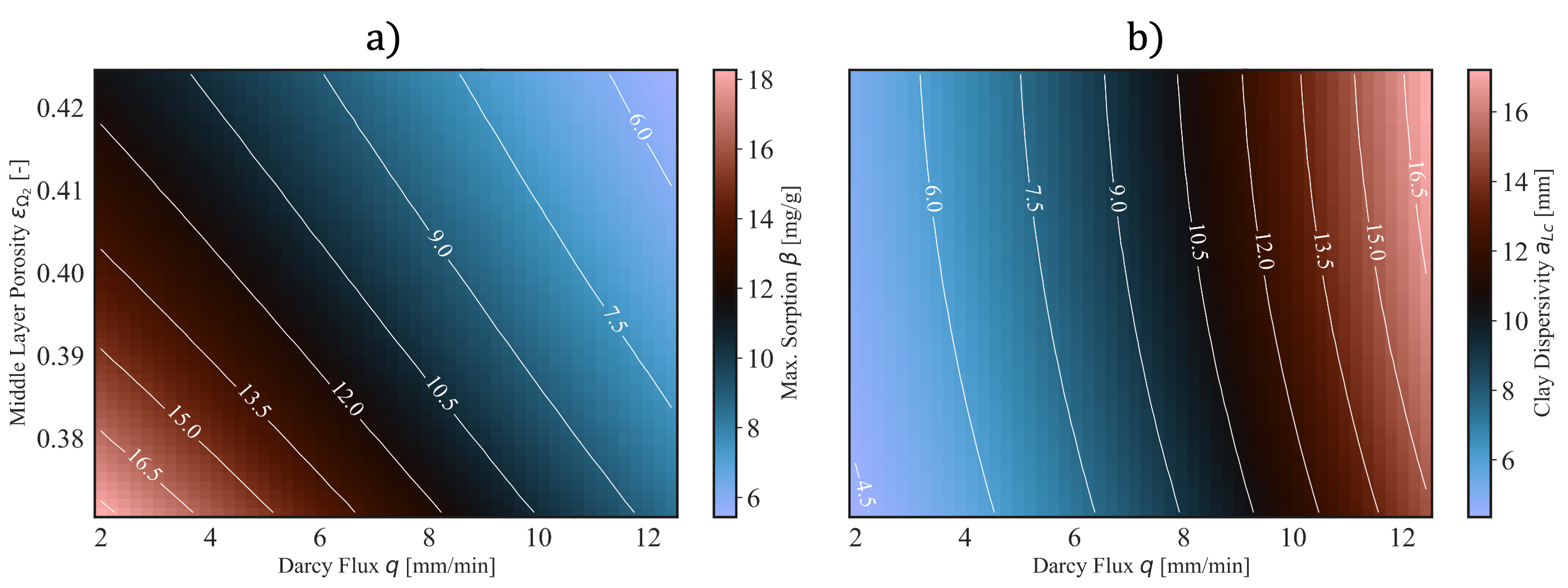}
    \caption{\textbf{Learned relationships between physical parameters.} The 2D heatmaps visualize the mean of the estimated physical parameters as a continuous function of the operational design space ($\varepsilon_{\Omega_2}, q$) at a fixed $C_0=100$ mg/L. \textbf{a)} The manifold for maximum sorption capacity, $\beta(\varepsilon_{\Omega_2}, q)$. \textbf{b)} The manifold for clay layer dispersivity, $a_{Lc}(\varepsilon_{\Omega_2}, q)$.}
    \label{fig:ch4_constitutive_heatmaps}
\end{figure}

We recall that the primary objective of the inverse framework is to learn continuous functional mappings between the operational design variables $\boldsymbol{\lambda}$ and the underlying physical parameters $\boldsymbol{\theta}$. To analyze these relationships, we evaluate the trained Parameter Network ($\mathcal{N}_{inv}$) over a two-dimensional slice of the operational design space defined by the middle layer porosity ($\varepsilon_{\Omega_2}$) and Darcy flux ($q$). This enables us to visualize the inferred dependencies as continuous parameter manifolds. Figure~\ref{fig:ch4_constitutive_heatmaps} depicts these manifolds for the two key physical quantities, i.e., the maximum sorption capacity ($\beta$) and the clay layer dispersivity ($a_{Lc}$), while holding the inlet concentration fixed at a baseline value $C_0=100$ mg/L.

The inferred manifold for sorption capacity, $\beta(\varepsilon_{\Omega_2}, q)$ (Figure~\ref{fig:ch4_constitutive_heatmaps}a), reveals two distinct and physically consistent trends. Along the vertical axis (corresponding to porosity), $\beta$ decreases as $\varepsilon_{\Omega_2}$ increases, reflecting the reduction in clay content and, consequently, the decrease in available sorption sites. The value of $\beta$ is then seen to decrease as Darcy flux increases,  indicating that shorter residence times at higher velocities limit access to intra-particle pore space and reduce the effective sorption capacity.

The manifold for clay dispersivity, $a_{Lc}(\varepsilon_{\Omega_2}, q)$ (Figure~\ref{fig:ch4_constitutive_heatmaps}b), reveals a systematic increase with Darcy flux, indicating that the inferred dispersivity behaves as an effective, flow-dependent parameter rather than a purely geometric constant. While classical formulations assume a linear relationship between dispersion and velocity, the observed trend suggests a nonlinear dependence, potentially reflecting unresolved transport mechanisms. In particular, rate-limited mass transfer processes (such as, e.g., intra-particle diffusion or non-equilibrium sorption) become more pronounced at higher flow rates and can markedly contribute to additional front spreading. In the absence of explicit kinetic terms in the governing ADR model, these effects are implicitly embedded into the inferred dispersivity values. Overall, these results suggest that the CoDe-BPINN framework identifies physically meaningful effective parameterizations that reconcile the adopted governing equations with the complex transport behavior observed in the experimental system.

\subsection{Local Sensitivity Analysis via Automatic Differentiation}
\label{sec:sensitivity}
\begin{figure}[h!]
    \centering
    \includegraphics[width=\textwidth]{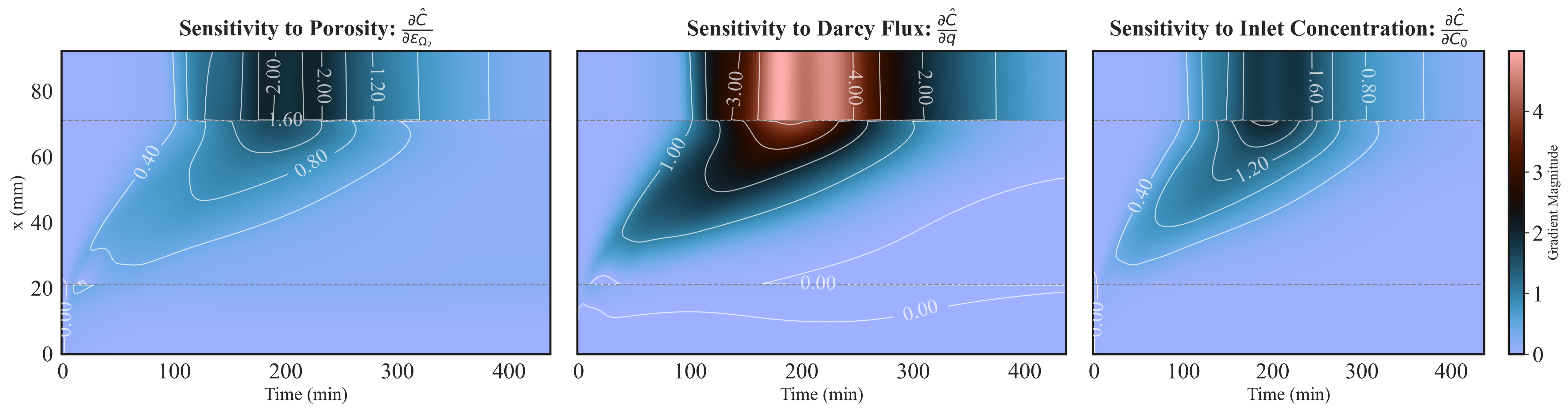}
    \caption{\textbf{Spatiotemporal Local Sensitivity Fields.} The maps illustrate the gradient of the mean predicted concentration, $\mathbb{E}[\hat{C}(\mathbf{x}; \boldsymbol{\lambda})]$, with respect to each of the normalized operational parameters $\boldsymbol{\lambda} = [\varepsilon_{\Omega_2}, q, C_0]$. The analysis is performed at the baseline condition ($\varepsilon_{\Omega_2}=0.371, q=1.99$ mm/min, $C_0=100$ mg/L). A positive sensitivity means that increasing the parameter increases concentration at that point, zero sensitivity indicates regions where the parameter has no influence.}
    \label{fig:ch4_sensitivity_maps}
\end{figure}

A distinct advantage of the CoDe-BPINN framework over conventional numerical solvers is its end-to-end differentiability. In traditional settings, Global or Local Sensitivity Analyses (GSA/LSA) typically require multiple forward simulations, leading to significant computational cost . In contrast, our trained surrogate model can be treated as a differentiable operator, enabling direct evaluation of local sensitivity indices, $\mathcal{S}_{\lambda} = \partial \hat{C} / \partial \boldsymbol{\lambda}$, through a single backward pass using Automatic Differentiation (AD). The ensuing sensitivity fields (see Figure~\ref{fig:ch4_sensitivity_maps}) provide detailed insight into the spatiotemporal regions where the system response is most sensitive to variations in physical and operational parameters.

Considering the impact of operational parameters on the concentration field, the sensitivity analysis reveals distinct and physically consistent patterns across the domain. We note that sensitivity tends to increase within the column length dimension, reflecting the accumulated influence of parameters across the transport path.

Sensitivity with respect to Porosity (as expressed through $\partial \hat{C}/\partial \varepsilon_{\Omega_2}$) is negligible within the inlet sand layer ($\Omega_1$), while becoming pronounced as the solute enters the reactive clay layer ($\Omega_2$). It further intensifies downstream in the outlet sand layer ($\Omega_3$). The predominantly positive gradient indicates that increasing porosity leads to earlier breakthrough. This behavior is physically consistent, as higher porosity corresponds to a reduced clay fraction and thus a lower density of sorption sites, diminishing the overall retardation capacity.

Sensitivity with respect to Darcy Flux (as expressed through $\partial \hat{C}/\partial q$)) exhibits the largest magnitude among the considered parameters and closely follows the propagation of the concentration front. A strong positive signal ahead of the front indicates that increasing the flow rate accelerates solute transport, leading to earlier arrival times. 

Finally, although the magnitude of sensitivity with respect to Inlet Concentration ($\partial \hat{C}/\partial C_0$)) is slightly lower than that associated with porosity, it is seen to form a localized \textit{plume} concentrated within the mass transfer zone of the reactive layer. This pattern reflects the role of inlet concentration in modulating sorption. In this context, higher values of $C_0$ accelerate saturation of available sorption sites and lead to earlier breakthrough.

This capability to evaluate sensitivity fields on demand across the entire design space is a key advantage of the proposed framework. It enables rapid assessment of the way variations in operational parameters propagate through the system in space and time, thus providing actionable insights into system behavior. Such analyses would be computationally prohibitive using traditional ensemble-based approaches, which require repeated forward simulations.

\section{Conclusions}
\label{sec:ch4_conclusion}

This study introduced a comprehensive Bayesian computational framework (termed CoDe-BPINN) that recasts the inverse problem of reactive transport as a coupled, end-to-end differentiable learning task. By integrating a domain-decomposed PINN solver with a multivariate parameter inference network, the proposed approach enables the reconstruction of contaminant transport dynamics in heterogeneous porous media from sparse boundary observations. The framework provides a robust methodology for jointly learning an accurate forward surrogate model and inferring the underlying constitutive relationships, even in data-limited settings. Our work leads to the following major conclusions.

\begin{enumerate}[label=(\roman*)]
    \item The proposed solver architecture, trained with a curriculum-based and physics-informed loss weighting, yields accurate reconstruction of the solution manifold across diverse experimental conditions. It captures sharp gradients at material interfaces and reproduces key features of the observed experimental breakthrough curves.
    
    \item The framework identifies multi-dimensional dependencies between the governing physical parameters and operational conditions. Notably, it reveals that the effective sorption capacity ($\beta$) decreases with increasing flow rate and porosity, reflecting the combined influence of kinetic limitations and reduced adsorbent mass.

    \item The multivariate parameter network captures the intrinsic structure of parameter interactions, including a strong negative correlation between sorption affinity ($\alpha$) and capacity ($\beta$), thereby providing a coeherent quantification of parameter uncertainty across the multidimensional design space.

    \item The framework yields predictive uncertainty in line with the uncertainty of the observations. Through Monte Carlo analysis and comparison with a reference FDM solver, it is shown that uncertainty remains low in data-constrained regions while being appropriately propagated into unobserved spatiotemporal domains, supporting the reliability of the model for predictive applications.
\end{enumerate}

Future developments may extend this methodology to higher-dimensional parameter spaces and enable real-time data assimilation for adaptive process monitoring and control.

\section*{Declarations}
\label{section_declaration}

\textit{Funding.} The authors acknowledge financial support from the European Union’s Horizon 2020 research and innovation programme under the Marie Skłodowska-Curie grant agreement No 956384\\\\
\textit{Code availability.} Codes used in this paper will be available in the following github repository \url{https://github.com/MiladPnh/Modulated-BPINNs.git}\\\\
\textit{Author Contributions.}
All authors came up with the research idea, M.P. and G.P designed the computation experiments, and M.P. implemented the code. G.P., A.G, and M.R. supervised the project. M.P. wrote the first draft, all participated in reviewing the manuscript.
\textit{Competing Interests.}
The authors have no competing interests to declare that are relevant to the content of this article.

\newpage
\bibliography{References}

@article{ceriotti2019double,
  title={A double-continuum transport model for segregated porous media: Derivation and sensitivity analysis-driven calibration},
  author={Ceriotti, G. and Russian, A. and Bolster, D. and Porta, G.},
  journal={Advances in Water Resources},
  volume={128},
  pages={206--217},
  year={2019},
  publisher={Elsevier}
}

@article{taylor2026global,
  title={Global sensitivity of laboratory-scale groundwater reactive transport simulations for sorption and degradation: Breakthrough dynamics across P{\'e}clet--Damk{\"o}hler regimes},
  author={Taylor, W. and Herman, J. and Morales, V.},
  journal={Journal of Contaminant Hydrology},
  pages={105043},
  year={2026},
  publisher={Elsevier}
}

@book{hairer1996solving2,
  author    = {Hairer, E. and Wanner, G.},
  title     = {Solving Ordinary Differential Equations II:
               Stiff and Differential-Algebraic Problems},
  edition   = {2},
  publisher = {Springer},
  year      = {1996},
  series    = {Springer Series in Computational Mathematics},
  volume    = {14},
  doi       = {10.1007/978-3-642-05221-7}
}

@book{bear2013dynamics,
  title={Dynamics of Fluids in Porous Media},
  author={Bear, J.},
  isbn={9780486656755},
  lccn={lc87034940},
  series={Dover Civil and Mechanical Engineering Series},
  url={https://books.google.it/books?id=lurrmlFGhTEC},
  year={1988},
  publisher={Dover}
}

@book{fetter1999contaminant,
  title={Contaminant hydrogeology},
  author={Fetter, C. W. and Boving, T. B. and Kreamer, D. K.},
  volume={1138},
  year={1999},
  publisher={Prentice hall Upper Saddle River, NJ}
}

@book{schiesser2012numerical,
  title={The numerical method of lines: integration of partial differential equations},
  author={Schiesser, W. E.},
  year={2012},
  publisher={Elsevier}
}

@book{rubin2003applied,
  title={Applied Stochastic Hydrogeology},
  author={Rubin, Y.},
  isbn={9780195138047},
  lccn={2002003663},
  series={OUP E-Books},
  url={https://books.google.it/books?id=CqLmCwAAQBAJ},
  year={2003},
  publisher={Oxford University Press}
}

@article{foo2010insights,
  title={Insights into the modeling of adsorption isotherm systems},
  author={Foo, K. Y. and Hameed, B. H.},
  journal={Chemical engineering journal},
  volume={156},
  number={1},
  pages={2--10},
  year={2010},
  publisher={Elsevier},
  url={https://doi.org/10.1016/j.cej.2009.09.013},
  doi={10.1016/j.cej.2009.09.013}
}

@article{limousin2007sorption,
title = {Sorption isotherms: A review on physical bases, modeling and measurement},
journal = {Applied Geochemistry},
volume = {22},
number = {2},
pages = {249-275},
year = {2007},
issn = {0883-2927},
doi = {https://doi.org/10.1016/j.apgeochem.2006.09.010},
url = {https://www.sciencedirect.com/science/article/pii/S0883292706002629},
author = {Limousin, G. and Gaudet, J. and Charlet, L. and Szenknect, S. and Barthès, V. and Krimissa, M.}
}

@article{worch2008fixed,
    author = {Worch, E.},
    title = {Fixed-bed adsorption in drinking water treatment: a critical review on models and parameter estimation},
    journal = {Journal of Water Supply: Research and Technology-Aqua},
    publisher={IWA Publishing},
    volume = {57},
    number = {3},
    pages = {171-183},
    year = {2008},
    month = {05},
    issn = {0003-7214},
    doi = {10.2166/aqua.2008.100},
    url = {https://doi.org/10.2166/aqua.2008.100},
    eprint = {https://iwaponline.com/aqua/article-pdf/57/3/171/401221/171.pdf},
}

@book{tarantola2005inverse,
author = {Tarantola, A.},
title = {Inverse Problem Theory and Methods for Model Parameter Estimation},
publisher = {Society for Industrial and Applied Mathematics},
year = {2005},
doi = {10.1137/1.9780898717921},
address = {},
edition   = {},
URL = {https://epubs.siam.org/doi/abs/10.1137/1.9780898717921},
eprint = {https://epubs.siam.org/doi/pdf/10.1137/1.9780898717921}
}

@article{beven2001equifinality,
title = {Equifinality, data assimilation, and uncertainty estimation in mechanistic modelling of complex environmental systems using the GLUE methodology},
journal = {Journal of Hydrology},
volume = {249},
number = {1},
pages = {11-29},
year = {2001},
issn = {0022-1694},
doi = {https://doi.org/10.1016/S0022-1694(01)00421-8},
url = {https://www.sciencedirect.com/science/article/pii/S0022169401004218},
author = {Beven, K. and Freer, J.}
}

@article{carrera2005inverse,
  title={Inverse problem in hydrogeology},
  author={Carrera, J. and Alcolea, A. and Medina, A. and Hidalgo, J. and Slooten, L. J.},
  journal={Hydrogeology journal},
  volume={13},
  number={1},
  pages={206--222},
  year={2005},
  publisher={Springer},
  url={https://doi.org/10.1007/s10040-004-0404-7},
  doi={10.1007/s10040-004-0404-7}
}

@book{scheidt2018quantifying,
publisher = {American Geophysical Union (AGU)},
isbn = {9781119325888},
title = {Front Matter},
booktitle = {Quantifying Uncertainty in Subsurface Systems},
author={Scheidt, C. and Li, L. and Caers, J.},
chapter = {},
pages = {i-xi},
doi = {https://doi.org/10.1002/9781119325888.fmatter},
url = {https://agupubs.onlinelibrary.wiley.com/doi/abs/10.1002/9781119325888.fmatter},
eprint = {https://agupubs.onlinelibrary.wiley.com/doi/pdf/10.1002/9781119325888.fmatter},
year={2018}}

@article{tartakovsky2013assessment, title={Assessment and management of risk in subsurface hydrology: A review and perspective}, volume={51}, ISSN={0309-1708}, url={http://dx.doi.org/10.1016/j.advwatres.2012.04.007}, DOI={10.1016/j.advwatres.2012.04.007}, journal={Advances in Water Resources}, publisher={Elsevier BV}, author={Tartakovsky, D. M.}, year={2013}, month=jan, pages={247–260}}

@misc{li2020fourier,
      title={Fourier Neural Operator for Parametric Partial Differential Equations}, 
      author={Li, Z. and Kovachki, N. and Azizzadenesheli, K. and Liu, B. and Bhattacharya, K. and Stuart, A. and Anandkumar, A.},
      year={2021},
      eprint={2010.08895},
      archivePrefix={arXiv},
      primaryClass={cs.LG},
      url={https://arxiv.org/abs/2010.08895}, 
}

@article{lu2021learning, title={Learning nonlinear operators via DeepONet based on the universal approximation theorem of operators}, volume={3}, ISSN={2522-5839}, url={http://dx.doi.org/10.1038/s42256-021-00302-5}, DOI={10.1038/s42256-021-00302-5}, number={3}, journal={Nature Machine Intelligence}, publisher={Springer Science and Business Media LLC}, author={Lu, L. and Jin, P. and Pang, G. and Zhang, Z. and Karniadakis, G. E.}, year={2021}, month=mar, pages={218–229} }

@article{luo2025physics,
  title={Physics-informed neural networks for PDE problems: a comprehensive review},
  author={Luo, K. and Zhao, J. and Wang, Y. and Li, J. and Wen, J. and Liang, J. and Soekmadji, H. and Liao, S.},
  journal={Artificial Intelligence Review},
  volume={58},
  number={10},
  pages={1--43},
  year={2025},
  publisher={Springer},
  doi={https://doi.org/10.1007/s10462-025-11322-7},
  url={https://link.springer.com/article/10.1007/s10462-025-11322-7#citeas}
}

@article{panahi2025modeling, title={Modeling parametric uncertainty in PDEs models via Physics-Informed Neural Networks}, volume={195}, ISSN={0309-1708},
url={http://dx.doi.org/10.1016/j.advwatres.2024.104870},
DOI={10.1016/j.advwatres.2024.104870},
journal={Advances in Water Resources},
publisher={Elsevier BV},
author={Panahi, M. and Porta, G. M. and Riva, M. and Guadagnini, A.}, 
year={2025}, 
pages={104870}}

@article{panahi2026pids,
    author = {Panahi, M. and Porta, G. M. and Riva, M. and Guadagnini, A.},
    title = {Physics informed differentiable solvers for learning parametric solution manifolds in heterogeneous physical systems},
    journal = {PNAS Nexus},
    pages = {pgag195},
    year = {2026},
    month = {06},
    issn = {2752-6542},
    doi = {10.1093/pnasnexus/pgag195},
    url = {https://doi.org/10.1093/pnasnexus/pgag195},
    eprint = {https://academic.oup.com/pnasnexus/advance-article-pdf/doi/10.1093/pnasnexus/pgag195/68440976/pgag195.pdf},
}

@article{SAdelloca,
author = {Dell’Oca, A.},
title = {Sensitivity Analysis: An Operational Picture},
journal = {Water Resources Research},
volume = {59},
number = {6},
pages = {e2022WR033780},
doi = {https://doi.org/10.1029/2022WR033780},
url = {https://agupubs.onlinelibrary.wiley.com/doi/abs/10.1029/2022WR033780},
eprint = {https://agupubs.onlinelibrary.wiley.com/doi/pdf/10.1029/2022WR033780},
note = {e2022WR033780 2022WR033780},
year = {2023}
}

@article{khan2026removal,
  title={Experimental and Modeling Investigations for High-Efficiency Removal of Polar Iodinated Contrast Media from Water through Engineered Clays},
  author={Khan, A. U. and Riva, M. and Porta, G. and Mauri, E. and Moscatelli, D. and Guadagnini, A.},
  journal={Chemical Engineering Journal Advances},
   doi={10.1016/j.ceja.2026.101351},
  pages={101351},
  year={2026},
  publisher={Elsevier}
}

@article{baydin2018automatic,
  title={Automatic differentiation in machine learning: a survey},
  author={Baydin, A. G. and Pearlmutter, B. A. and Radul, A. A. and Siskind, J. M.},
  journal={Journal of machine learning research},
  volume={18},
  number={153},
  pages={1--43},
  year={2018},
  url={http://jmlr.org/papers/v18/17-468.html}
}

@inproceedings{NEURIPS2018_DP,
 author = {de Avila Belbute-Peres, F. and Smith, K. and Allen, K. and Tenenbaum, J. and Kolter, J. Z.},
 booktitle = {Advances in Neural Information Processing Systems},
 editor = {S. Bengio and H. Wallach and H. Larochelle and K. Grauman and N. Cesa-Bianchi and R. Garnett},
 pages = {},
 publisher = {Curran Associates, Inc.},
 title = {End-to-End Differentiable Physics for Learning and Control},
 url = {https://proceedings.neurips.cc/paper_files/paper/2018/file/842424a1d0595b76ec4fa03c46e8d755-Paper.pdf},
 volume = {31},
 year = {2018}
}

@misc{ha2016hypernetworks,
      title={HyperNetworks}, 
      author={Ha, D. and Dai, A. and Le, Q. V.},
      year={2016},
      eprint={1609.09106},
      archivePrefix={arXiv},
      primaryClass={cs.LG},
      url={https://arxiv.org/abs/1609.09106}, 
}

@article{CHEN2022110996,
title = {Meta-MgNet: Meta multigrid networks for solving parameterized partial differential equations},
journal = {Journal of Computational Physics},
volume = {455},
pages = {110996},
year = {2022},
issn = {0021-9991},
doi = {https://doi.org/10.1016/j.jcp.2022.110996},
url = {https://www.sciencedirect.com/science/article/pii/S0021999122000584},
author = {Chen, Y. and Dong, B. and Xu, J.}
}

@misc{nichol2018reptile,
      title={On First-Order Meta-Learning Algorithms}, 
      author={Nichol, A. and Achiam, J. and Schulman, J.},
      year={2018},
      eprint={1803.02999},
      archivePrefix={arXiv},
      primaryClass={cs.LG},
      url={https://arxiv.org/abs/1803.02999}, 
}

@article{ANAGNOSTOPOULOS2026107983,
title = {Learning in PINNs: Phase transition, diffusion equilibrium, and generalization},
journal = {Neural Networks},
volume = {193},
pages = {107983},
year = {2026},
issn = {0893-6080},
doi = {https://doi.org/10.1016/j.neunet.2025.107983},
url = {https://www.sciencedirect.com/science/article/pii/S0893608025008640},
author = {Anagnostopoulos, S. J. and Toscano, J. D. and Stergiopulos, N. and Karniadakis, G. E.}
}

@article{tancik2020fourier,
  title={Fourier features let networks learn high frequency functions in low dimensional domains},
  author={Tancik, M. and Srinivasan, P. and Mildenhall, B. and Fridovich-Keil, S. and Raghavan, N. and Singhal, U. and Ramamoorthi, R. and Barron, J. and Ng, R.},
  journal={Advances in neural information processing systems},
  volume={33},
  pages={7537--7547},
  year={2020},
  url={https://papers.neurips.cc/paper_files/paper/2020/file/55053683268957697aa39fba6f231c68-Paper.pdf}
}

@misc{thuerey2021physics,
      title={Physics-based Deep Learning}, 
      author={Thuerey, N. and Holzschuh, B. and Holl, P. and Kohl, G. and Lino, M. and Liu, Q. and Schnell, P. and Trost, F.},
      year={2025},
      eprint={2109.05237},
      archivePrefix={arXiv},
      primaryClass={cs.LG},
      url={https://arxiv.org/abs/2109.05237}, 
}

@misc{flores2025improved,
      title={Improved Uncertainty Quantification in Physics-Informed Neural Networks Using Error Bounds and Solution Bundles}, 
      author={Flores, P. and Graf, O. and Protopapas, P. and Pichara, K.},
      year={2025},
      eprint={2505.06459},
      archivePrefix={arXiv},
      primaryClass={cs.LG},
      url={https://arxiv.org/abs/2505.06459}, 
}

@article{wang2021understanding, title={Understanding and Mitigating Gradient Flow Pathologies in Physics-Informed Neural Networks}, volume={43}, ISSN={1095-7197}, url={http://dx.doi.org/10.1137/20m1318043}, DOI={10.1137/20m1318043}, number={5}, journal={SIAM Journal on Scientific Computing}, publisher={Society for Industrial & Applied Mathematics (SIAM)}, author={Wang, S. and Teng, Y. and Perdikaris, P.}, year={2021}, month=jan, pages={A3055–A3081} }

@article{yang2021b,
title = {B-PINNs: Bayesian physics-informed neural networks for forward and inverse PDE problems with noisy data},
journal = {Journal of Computational Physics},
volume = {425},
pages = {109913},
year = {2021},
issn = {0021-9991},
doi = {https://doi.org/10.1016/j.jcp.2020.109913},
url = {https://www.sciencedirect.com/science/article/pii/S0021999120306872},
author = {Yang, L. and Meng, X. and Karniadakis, G. E.}
}

@article{meng2021multi,
title = {Multi-fidelity Bayesian neural networks: Algorithms and applications},
journal = {Journal of Computational Physics},
volume = {438},
pages = {110361},
year = {2021},
issn = {0021-9991},
doi = {https://doi.org/10.1016/j.jcp.2021.110361},
url = {https://www.sciencedirect.com/science/article/pii/S0021999121002564},
author = {Meng, X. and Babaee, H. and Karniadakis, G. E.}
}

@misc{lavin2021simulation,
      title={Simulation Intelligence: Towards a New Generation of Scientific Methods}, 
      author={Lavin, A. and Krakauer, D. and Zenil, H. and Gottschlich, J. and Mattson, T. and Brehmer, J. and Anandkumar, A. and Choudry, S. and Rocki, K. and Baydin, A. G. and Prunkl, C. and Paige, B. and Isayev, O. and Peterson, E. and McMahon, P. L. and Macke, J. and Cranmer, K. and Zhang, J. and Wainwright, H. and Hanuka, A. and Veloso, M. and Assefa, S. and Zheng, S. and Pfeffer, A.},
      year={2022},
      eprint={2112.03235},
      archivePrefix={arXiv},
      primaryClass={cs.AI},
      url={https://arxiv.org/abs/2112.03235}, 
}

@InProceedings{rahaman2019spectral,
  title = 	 {On the Spectral Bias of Neural Networks},
  author =       {Rahaman, N. and Baratin, A. and Arpit, D. and Draxler, F. and Lin, M. and Hamprecht, F. and Bengio, Y. and Courville, A.},
  booktitle = 	 {Proceedings of the 36th International Conference on Machine Learning},
  pages = 	 {5301--5310},
  year = 	 {2019},
  editor = 	 {Chaudhuri, Kamalika and Salakhutdinov, Ruslan},
  volume = 	 {97},
  series = 	 {Proceedings of Machine Learning Research},
  month = 	 {09--15 Jun},
  publisher =    {PMLR},
  url = 	 {https://proceedings.mlr.press/v97/rahaman19a.html}
}

@article{krishnapriyan2021characterizing,
  title={Characterizing possible failure modes in physics-informed neural networks},
  author={Krishnapriyan, A. and Gholami, A. and Zhe, S. and Kirby, R. and Mahoney, M. W.},
  journal={Advances in neural information processing systems},
  volume={34},
  pages={26548--26560},
  year={2021},
  url = {https://proceedings.neurips.cc/paper/2021/file/df438e5206f31600e6ae4af72f2725f1-Paper.pdf}
}

@inbook{gupta2006model,
author = {Gupta, H. V. and Beven, K. J. and Wagener, T.},
publisher = {John Wiley \& Sons, Ltd},
isbn = {9780470848944},
title = {Model Calibration and Uncertainty Estimation},
booktitle = {Encyclopedia of Hydrological Sciences},
chapter = {131},
pages = {},
doi = {https://doi.org/10.1002/0470848944.hsa138},
url = {https://onlinelibrary.wiley.com/doi/abs/10.1002/0470848944.hsa138},
eprint = {https://onlinelibrary.wiley.com/doi/pdf/10.1002/0470848944.hsa138},
year = {2006}
}

@article{dell2019solute, title={Solute transport in random composite media with uncertain dispersivities}, volume={128}, ISSN={0309-1708}, url={http://dx.doi.org/10.1016/j.advwatres.2019.04.005}, DOI={10.1016/j.advwatres.2019.04.005}, journal={Advances in Water Resources}, publisher={Elsevier BV}, author={Dell’Oca, A. and Riva, M. and Ackerer, P. and Guadagnini, A.}, year={2019}, month=jun, pages={48–58} }

@article{asher2015review, title={A review of surrogate models and their application to groundwater modeling}, volume={51}, ISSN={1944-7973}, url={http://dx.doi.org/10.1002/2015wr016967}, DOI={10.1002/2015wr016967}, number={8}, journal={Water Resources Research}, publisher={American Geophysical Union (AGU)}, author={Asher, M. J. and Croke, B. F. W. and Jakeman, A. J. and Peeters, L. J. M.}, year={2015}, month=aug, pages={5957–5973} }

@article{Maina2018, title={Uncertainty Quantification and Global Sensitivity Analysis of Subsurface Flow Parameters to Gravimetric Variations During Pumping Tests in Unconfined Aquifers}, volume={54}, ISSN={1944-7973}, url={http://dx.doi.org/10.1002/2017wr021655}, DOI={10.1002/2017wr021655}, number={1}, journal={Water Resources Research}, publisher={American Geophysical Union (AGU)}, author={Maina, F. Z. and Guadagnini, A.}, year={2018}, month=jan, pages={501–518} }

@article{CHEN2021110666, title={Physics-informed machine learning for reduced-order modeling of nonlinear problems}, volume={446}, ISSN={0021-9991}, url={http://dx.doi.org/10.1016/j.jcp.2021.110666}, DOI={10.1016/j.jcp.2021.110666}, journal={Journal of Computational Physics}, publisher={Elsevier BV}, author={Chen, W. and Wang, Q. and Hesthaven, J. S. and Zhang, C.}, year={2021}, month=dec, pages={110666} }

@article{jagtap2020conservative,
title = {Conservative physics-informed neural networks on discrete domains for conservation laws: Applications to forward and inverse problems},
journal = {Computer Methods in Applied Mechanics and Engineering},
volume = {365},
pages = {113028},
year = {2020},
issn = {0045-7825},
doi = {https://doi.org/10.1016/j.cma.2020.113028},
url = {https://www.sciencedirect.com/science/article/pii/S0045782520302127},
author = {Jagtap, A. D. and Kharazmi, E. and Karniadakis, G. E.}
}

@article{lagaris1998artificial, title={Artificial neural networks for solving ordinary and partial differential equations}, volume={9}, ISSN={1045-9227}, url={http://dx.doi.org/10.1109/72.712178}, DOI={10.1109/72.712178}, number={5}, journal={IEEE Transactions on Neural Networks}, publisher={Institute of Electrical and Electronics Engineers (IEEE)}, author={Lagaris, I. and Likas, A. and Fotiadis, D.}, year={1998}, pages={987–1000} }

@inbook{fletcher2000practical,
author = {Fletcher, R.},
publisher = {John Wiley \& Sons, Ltd},
isbn = {9781118723203},
title = {The Theory of Constrained Optimization},
booktitle = {Practical Methods of Optimization},
chapter = {9},
pages = {195-228},
doi = {https://doi.org/10.1002/9781118723203.ch9},
url = {https://onlinelibrary.wiley.com/doi/abs/10.1002/9781118723203.ch9},
eprint = {https://onlinelibrary.wiley.com/doi/pdf/10.1002/9781118723203.ch9},
year = {2000}
}

@misc{raissi2023open,
      title={Open Problems in Applied Deep Learning}, 
      author={Raissi, M.},
      year={2023},
      eprint={2301.11316},
      archivePrefix={arXiv},
      primaryClass={cs.LG},
      url={https://arxiv.org/abs/2301.11316}, 
}

@article{chen2020physics,
  title={Physics-informed neural networks for inverse problems in nano-optics and metamaterials},
  author={Chen, Y. and Lu, L. and Karniadakis, G. E. and Dal Negro, L.},
  journal={Optics express},
  volume={28},
  number={8},
  pages={11618--11633},
  year={2020},
  publisher={OSA}
}

@article{zhu2019physics, title={Physics-constrained deep learning for high-dimensional surrogate modeling and uncertainty quantification without labeled data}, volume={394}, ISSN={0021-9991}, url={http://dx.doi.org/10.1016/j.jcp.2019.05.024}, DOI={10.1016/j.jcp.2019.05.024}, journal={Journal of Computational Physics}, publisher={Elsevier BV}, author={Zhu, Y. and Zabaras, N. and Koutsourelakis, P. and Perdikaris, P.}, year={2019}, month=oct, pages={56–81} }

@article{wang2022and, title={When and why PINNs fail to train: A neural tangent kernel perspective}, volume={449}, ISSN={0021-9991}, url={http://dx.doi.org/10.1016/j.jcp.2021.110768}, DOI={10.1016/j.jcp.2021.110768}, journal={Journal of Computational Physics}, publisher={Elsevier BV}, author={Wang, S. and Yu, X. and Perdikaris, P.}, year={2022}, month=jan, pages={110768} }

@article{sun2020surrogate, title={Surrogate modeling for fluid flows based on physics-constrained deep learning without simulation data}, volume={361}, ISSN={0045-7825}, url={http://dx.doi.org/10.1016/j.cma.2019.112732}, DOI={10.1016/j.cma.2019.112732}, journal={Computer Methods in Applied Mechanics and Engineering}, publisher={Elsevier BV}, author={Sun, L. and Gao, H. and Pan, S. and Wang, J.}, year={2020}, month=apr, pages={112732} }

@article{kovachki2021neural,
  author       = {Kovachki, N. B. and Li, Z. and Liu, B. and Azizzadenesheli, K. and Bhattacharya, K. and Stuart, A. M. and Anandkumar, A.},
  title        = {Neural Operator: Learning Maps Between Function Spaces},
  journal      = {CoRR},
  volume       = {abs/2108.08481},
  year         = {2021},
  url          = {https://arxiv.org/abs/2108.08481},
  eprinttype    = {arXiv},
  eprint       = {2108.08481},
  bibsource    = {dblp computer science bibliography, https://dblp.org}
}

@misc{innes2019differentiable,
      title={A Differentiable Programming System to Bridge Machine Learning and Scientific Computing}, 
      author={Innes, M. and Edelman, A. and Fischer, K. and Rackauckas, C. and Saba, E. and Shah, V. B. and Tebbutt, W.},
      year={2019},
      eprint={1907.07587},
      archivePrefix={arXiv},
      primaryClass={cs.PL},
      url={https://arxiv.org/abs/1907.07587}, 
}

@article{lee2009comparative, title={A comparative study of uncertainty propagation methods for black-box-type problems}, volume={37}, ISSN={1615-1488}, url={http://dx.doi.org/10.1007/s00158-008-0234-7}, DOI={10.1007/s00158-008-0234-7}, number={3}, journal={Structural and Multidisciplinary Optimization}, publisher={Springer Science and Business Media LLC}, author={Lee, S. H. and Chen, W.}, year={2008}, month=may, pages={239–253} }

@article{zhang2019quantifying, title={Quantifying total uncertainty in physics-informed neural networks for solving forward and inverse stochastic problems}, volume={397}, ISSN={0021-9991}, url={http://dx.doi.org/10.1016/j.jcp.2019.07.048}, DOI={10.1016/j.jcp.2019.07.048}, journal={Journal of Computational Physics}, publisher={Elsevier BV}, author={Zhang, D. and Lu, L. and Guo, L. and Karniadakis, G. E.}, year={2019}, month=nov, pages={108850} }

@article{sun2020physics,
title = {Physics-constrained bayesian neural network for fluid flow reconstruction with sparse and noisy data},
journal = {Theoretical and Applied Mechanics Letters},
volume = {10},
number = {3},
pages = {161-169},
year = {2020},
issn = {2095-0349},
doi = {https://doi.org/10.1016/j.taml.2020.01.031},
url = {https://www.sciencedirect.com/science/article/pii/S2095034920300295},
author = {Sun, L. and Wang, J.}
}

@article{psaros2023uncertainty,
title = {Uncertainty quantification in scientific machine learning: Methods, metrics, and comparisons},
journal = {Journal of Computational Physics},
volume = {477},
pages = {111902},
year = {2023},
issn = {0021-9991},
doi = {https://doi.org/10.1016/j.jcp.2022.111902},
url = {https://www.sciencedirect.com/science/article/pii/S0021999122009652},
author = {Psaros, A. F. and Meng, X. and Zou, Z. and Guo, L. and Karniadakis, G. E.}
}

@article{JAGTAP2020113028,
title = {Conservative physics-informed neural networks on discrete domains for conservation laws: Applications to forward and inverse problems},
journal = {Computer Methods in Applied Mechanics and Engineering},
volume = {365},
pages = {113028},
year = {2020},
issn = {0045-7825},
doi = {https://doi.org/10.1016/j.cma.2020.113028},
url = {https://www.sciencedirect.com/science/article/pii/S0045782520302127},
author = {Jagtap, A. D. and Kharazmi, E. and Karniadakis, G. E.}
}

@misc{kingma2013auto,
      title={Auto-Encoding Variational Bayes}, 
      author={Kingma, D. P. and Welling, M.},
      year={2022},
      eprint={1312.6114},
      archivePrefix={arXiv},
      primaryClass={stat.ML},
      url={https://arxiv.org/abs/1312.6114}, 
}

@article{KHAN2024116506,
title = {In-silico mechanistic analysis of adsorption of Iodinated Contrast Media agents on graphene surface},
journal = {Ecotoxicology and Environmental Safety},
volume = {280},
pages = {116506},
year = {2024},
issn = {0147-6513},
doi = {https://doi.org/10.1016/j.ecoenv.2024.116506},
url = {https://www.sciencedirect.com/science/article/pii/S0147651324005827},
author = {Khan, A. U. and Porta, G. M. and Riva, M. and Guadagnini, A.}
}

@article{sengar2021occurrence,
title = {Fate and removal of iodinated X-ray contrast media in membrane bioreactor: Microbial dynamics and effects of different operational parameters},
journal = {Science of The Total Environment},
volume = {869},
pages = {161827},
year = {2023},
issn = {0048-9697},
doi = {https://doi.org/10.1016/j.scitotenv.2023.161827},
url = {https://www.sciencedirect.com/science/article/pii/S0048969723004424},
author = {Sengar, A. and Vijayanandan, A.}
}

@article{kormos2011occurrence,
author = {Kormos, J. L. and Schulz, M. and Ternes, T. A.},
title = {Occurrence of Iodinated X-ray Contrast Media and Their Biotransformation Products in the Urban Water Cycle},
journal = {Environmental Science \& Technology},
publisher={ACS Publications},
volume = {45},
number = {20},
pages = {8723-8732},
year = {2011},
doi = {10.1021/es2018187},
note ={PMID: 21877755},
URL = {https://doi.org/10.1021/es2018187},
eprint = { https://doi.org/10.1021/es2018187}}

@article{duirk2011formation,
author = {Duirk, S. E. and Lindell, C. and Cornelison, C. C. and Kormos, J. and Ternes, T. A. and Attene-Ramos, M. and Osiol, J. and Wagner, E. D. and Plewa, M. J. and Richardson, S. D.},
title = {Formation of Toxic Iodinated Disinfection By-Products from Compounds Used in Medical Imaging},
journal = {Environmental Science \& Technology},
volume = {45},
number = {16},
pages = {6845-6854},
year = {2011},
doi = {10.1021/es200983f},
note ={PMID: 21761849},
URL = {https://doi.org/10.1021/es200983f},
eprint = {https://doi.org/10.1021/es200983f}}

@article{patel2022comparison,
  title={Comparison of batch and fixed bed column adsorption: a critical review},
  author={Patel, H.},
  journal={International Journal of Environmental Science and Technology},
  volume={19},
  number={10},
  pages={10409--10426},
  year={2022},
  publisher={Springer},
  url={https://doi.org/10.1007/s13762-021-03492-y},
  doi={10.1007/s13762-021-03492-y}
}

@article{raissi2019physics, title={Physics-informed neural networks: A deep learning framework for solving forward and inverse problems involving nonlinear partial differential equations}, volume={378}, ISSN={0021-9991}, url={http://dx.doi.org/10.1016/j.jcp.2018.10.045}, DOI={10.1016/j.jcp.2018.10.045}, journal={Journal of Computational Physics}, publisher={Elsevier BV}, author={Raissi, M. and Perdikaris, P. and Karniadakis, G.}, year={2019}, month=feb, pages={686–707} }

@article{penwarden2023metalearning, title={A metalearning approach for Physics-Informed Neural Networks (PINNs): Application to parameterized PDEs}, volume={477}, ISSN={0021-9991}, url={http://dx.doi.org/10.1016/j.jcp.2023.111912}, DOI={10.1016/j.jcp.2023.111912}, journal={Journal of Computational Physics}, publisher={Elsevier BV}, author={Penwarden, M. and Zhe, S. and Narayan, A. and Kirby, R. M.}, year={2023}, month=mar, pages={111912} }

@misc{raissi2024physics,
      title={Physics-Informed Neural Networks and Extensions}, 
      author={Raissi, M. and Perdikaris, P. and Ahmadi, N. and Karniadakis, G. E.},
      year={2024},
      eprint={2408.16806},
      archivePrefix={arXiv},
      primaryClass={cs.LG},
      url={https://arxiv.org/abs/2408.16806}
}

\newpage
\appendix

\section*{Supporting information}
\setcounter{equation}{0}
\renewcommand{\theequation}{S\arabic{equation}}
\setcounter{figure}{0}
\renewcommand{\thefigure}{S\arabic{figure}}
\setcounter{table}{0}
\renewcommand{\thetable}{S\arabic{table}}

\begin{abstract}
This supplementary information file is divided into two sections:
\begin{enumerate}
    \item Description of the reference numerical solver, employed to verify our CoDe-BPINN results.
    \item  Discussion of hyperparameters and training strategy technical details.
\end{enumerate}
\end{abstract}

\section{Reference Numerical Framework: Method-of-Lines Finite Difference Solver}
\label{sec:ch4_fdm_framework}
To rigorously validate the results ($\mathcal{S}$) stemming from our approach, we rely on a high-fidelity reference solver based on the Method of Lines (MOL) \citep{schiesser2012numerical}. This approach discretizes the governing equation (Equation 4) in space while retaining the continuous time dynamics, thereby transforming the problem into a system of coupled Ordinary Differential Equations (ODEs).

\paragraph{Spatial Discretization.}
The spatial domain $\Omega = [0, L]$ is discretized using a uniform grid of $N_x$ nodes with spacing $\Delta x = L/(N_x - 1)$. The state vector $\mathbf{C}(t) \in \mathbb{R}^{N_x}$ represents the concentration at each grid point. To account for the advection-dominated transport regime, the advective flux is approximated through a first-order upwind scheme, which ensures numerical stability (monotonicity), albeit at the expense of introducing a mild numerical diffusion. The dispersive flux is discretized through a second-order central difference scheme. Heterogeneous material properties (namely porosity $\varepsilon_i$, dispersivity $a_{L,i}$, and retardation factor $R_i$) are assigned on node-wise basis according to the column stratification, sharp discontinuities being explicitly preserved at layer interfaces ($x=21$ mm and $x=71$ mm).

\paragraph{Boundary Conditions.}
The Dirichlet inlet condition is imposed directly as a state constraint, i.e., $C_0(t) = 1$. For the Neumann outlet condition ($\partial C/\partial x |_{x=L} = 0$), we employ a ghost node formulation to retain second-order accuracy. A fictitious node $C_{N+1}$ is introduced outside the domain and defined such that $C_{N+1} = C_{N-1}$. This leads to the following approximation for the second derivative associated with the dispersive term at the outlet node $N$:
\begin{equation}
\left. \frac{\partial^2 C}{\partial x^2} \right|_N \approx \frac{2C_{N-1} - 2C_N}{\Delta x^2}
\end{equation}
This formulation enforces zero dispersive flux at the boundary while allowing advective outflow.

\paragraph{Non-Linear Reaction and Time Integration.}
The non-linear Langmuir sorption introduces a concentration-dependent Retardation Factor $R_L(C)$. To address the ensuing stiffness of the system while preserving mass conservation, we employ an implicit time integration scheme. The system of ODEs, $\frac{d\mathbf{C}}{dt} = \mathbf{F}(t, \mathbf{C})$, is integrated using the Radau IIA method \citep{hairer1996solving2} (an implicit Runge-Kutta scheme of order 5), which is well suited to capturing the sharp transients associated with the propagation front. The Jacobian of the system is approximated via finite differences to facilitate Newton-Raphson convergence at each time step.

\paragraph{Uncertainty Propagation.}
In contrast to the Solver $\mathcal{S}$ within our CoDe-BPINN framework, which provides probabilistic results, the FDM solver operates on fixed parameter realizations and does not intrinsically quantify uncertainty. To obtain reference uncertainty estimates for comparison, we perform a Monte Carlo ensemble analysis. We draw $N_{MC}$ samples of the physical parameter vector $\boldsymbol{\theta}$ from the posterior distribution learned by the Parameter Network, $q(\boldsymbol{\theta}|\boldsymbol{\lambda})$, and execute the FDM solver for each of these realizations. The resulting ensemble statistics (namely, mean and standard deviation) serve as reference quantities (i.e., ground truth) to assess epistemic uncertainty estimates captured by the CoDe-BPINN.

\subsection{Numerical Verification of the Reference Solver}
\label{sec:ch4_fdm_reference}
To ensure reliability of the reference solution used for validation, the Method-of-Lines FDM solver was subject to a grid convergence analysis. The spatial domain was discretized using $N_x=200$ nodes (corresponding to a spatial resolution $\Delta x = 0.46$ mm), and time integration was performed using an adaptive Runge-Kutta method (Radau IIA) with absolute and relative tolerances of $10^{-6}$. Refining the spatial discretization to $N_x=500$ resulted in a relative change in the breakthrough curve below $0.1\%$, suggesting that the numerical solution is effectively mesh-independent and sufficiently accurate to be employed as benchmark for validation of the CoDe-BPINN results.

\subsection{Computational Efficiency and Inference  cost}
\label{sec:ch4_Computational}
A primary motivation for developing the CoDe-BPINN digital twin is to replace computationally expensive iterative solvers with near-instantaneous neural inference. The training phase, entailing joint optimization of the solver $\mathcal{S}$ and Parameter Inference Network $\mathcal{N}_{inv}$ over 150,000 epochs, required approximately 2 hours on a single NVIDIA RTX4090 GPU. While this upfront cost exceeds that of a single forward simulation using a Finite Difference Method (FDM) solver ($\approx$ 5 seconds), the advantage becomes evident at inference.
Once trained, the CoDe-BPINN acts as a fully parameterized surrogate model. It can generate a complete spatiotemporal solution field for a new set of operational design parameters in $\mathcal{O}(10^{-3})$ seconds, corresponding to a speed-up of approximately three orders of magnitude relative to the numerical solver. Furthermore, the CoDe-BPINN provides access to the full Jacobian of the solution with respect to input parameters via Automatic Differentiation (AD) \citep{baydin2018automatic}. This enables real-time sensitivity analysis and uncertainty propagation, a task that would otherwise require large ensembles of Monte Carlo simulations in a conventional FDM framework.

\begin{table}[!ht]
    \centering
    \caption{Hyperparameter configuration for CoDe-BPINN.}
    \label{tab:ch4_hyperparams_case1}
    \begin{tabular}{@{}ll}
        \toprule
        Parameter & Value \\
        \midrule
        \multicolumn{2}{l}{\textbf{Architecture (Solver $\mathcal{S}$)}} \\
        Number of sub-networks & 3 ($\mathcal{N}^{\Omega_1}, \mathcal{N}^{\Omega_2}, \mathcal{N}^{\Omega_3}$) \\
        Input dimension & $d_{\boldsymbol{\lambda}} +  d= 3 + 2$ \\
        Number of Bayesian Dense Layers & 2 \\
        Neurons per hidden layer & 100 \\
        Activation function & SiLU (hidden), Sigmoid (output) \\
        \multicolumn{2}{l}{\textbf{Architecture (Parameter Network $\mathcal{N}_{inv}$)}} \\
        Input dimension & $d_{\boldsymbol{\lambda}} = 3$ \\
        Hidden dimension & 6 \\
        Output dimension & 14 ($d_{\boldsymbol{\theta}} = 4$  Means + 10 Cholesky elements) \\
        Activation function & Tanh (hidden), Linear (output) \\
        \midrule 
        \multicolumn{2}{l}{\textbf{Input Embedding}} \\
        Input embedding type & Random Fourier Features \\
        Fourier feature scale & 1.0 \\
        Embedding dimension & 16 \\
        \midrule
        \multicolumn{2}{l}{\textbf{Optimizer}} \\
        Optimizer type & AdamW \\
        Weight decay & $10^{-4}$ \\
        Initial learning rate & $10^{-3}$ \\
        \midrule
        \multicolumn{2}{l}{\textbf{Learning Rate Schedule}} \\
        Warmup steps & 1,000 \\
        Decay schedule & Cosine annealing schedule \\
        Total Epochs & 150,000 \\
        \midrule
        \multicolumn{2}{l}{\textbf{Curriculum \& Loss Weighting}} \\
        Stage 1 (Geometric Init) & 2,000 epochs \\
        Stage 2 (Data Anchoring) & 6,000 epochs \\
        Stage 3 (Physics Discovery) & 142,000 epochs \\
        KL Annealing Start & Epoch 15,000 \\
        KL Weight Range & [$10^{-5}, 10^{-2}$] \\
        Loss Weights ($w_{data}, w_{pde}$) & (200.0, $5 \times 10^3$) \\
        Loss Weights ($w_{bc}, w_{ic}, w_{int}$) & ($8 \times 10^4$, $8 \times 10^4$, $8 \times 10^4$) \\
        \midrule
        \multicolumn{2}{l}{\textbf{Collocation Sampling}} \\
        Batch size (PDE) & $2^{14}$ \\
        Batch size (Data) & $2^{11}$ \\
        Batch size (BCs/IC) & $2^{14}$ \\
        Interface Points ($N_{int}$) & $2^{9}$ \\
        Sampling Method & Sobol Sequences (Scrambled) \\
        \bottomrule
    \end{tabular}
\end{table}

\section{Hyperparameter Configurations}
\label{sec:ch4_hyperparams}
Table \ref{tab:ch4_hyperparams_case1} lists the key elements of the network architecture, optimizer settings, and training hyperparameters used in this study.

\paragraph{Priors and Initialization.}
For the solver weights $\boldsymbol{\psi}$, we employ a standard isotropic Gaussian prior $N(0, 1)$, which regularizes the magnitude of the weights and promotes smooth solutions. The informative priors assigned to the physical parameters $\boldsymbol{\theta}$ (including first two moments associated with the Log-Normal distributions governing dispersivity, affinity, and capacity) are defined in detail in Section 3.5.3 and are implemented accordingly.

\paragraph{Training strategy.}
To enhance training stability and address stiffness of the loss landscape, we employ two complementary annealing scheduling strategies. The learning rate of the AdamW optimizer is managed by a cosine annealing schedule with an initial linear warmup phase. The learning rate increases from zero to a maximum of $10^{-3}$ over the first 1,000 epochs, facilitating robust exploration and mitigating poor initial convergence, and then smoothly decays towards a minimum value of $10^{-5}$ over the remainder of training to promote fine-tuning. Concurrently, the weight of the KL divergence term in Equation 13, $w_{kl}$, is progressively increased during training. It is initially set to a small value ($10^{-5}$) for the first 15,000 epochs, allowing the solver to flexibly fit observational data and boundary conditions. It is then linearly increased to a final value of $10^{-2}$, gradually strengthening the influence of the prior. This scheduling strategy promotes a transition from data-driven fitting to regularized inference, reducing epistemic uncertainty in the Bayesian solver $\mathcal{S}$ and forcing the Parameter Inference Network to account for the governing physical residuals, thereby preventing the solver from learning unphysical solution behavior.

\end{document}